\documentclass[zpreprint,zbstepj]{zeus_paper}
\usepackage[english]{babel}
\newcommand{\DO}{{D{\O}}\xspace}

\chardef\usc=95
\chardef\til=126
\catcode`\@=11 
\DeclareRobustCommand\xdotspace{\futurelet\@let@token\@xdotspace}
\def\@xdotspace{%
  \ifx\@let@token.\else
  \ifx\@let@token\bgroup.\else
  \ifx\@let@token\egroup.\else
  \ifx\@let@token\/.\else
  \ifx\@let@token\ .\else
  \ifx\@let@token~.\else
  \ifx\@let@token!.\else
  \ifx\@let@token,.\else
  \ifx\@let@token:.\else
  \ifx\@let@token;.\else
  \ifx\@let@token?.\else
  \ifx\@let@token/.\else
  \ifx\@let@token'.\else
  \ifx\@let@token).\else
  \ifx\@let@token-.\else
  \ifx\@let@token\@xobeysp.\else
  \ifx\@let@token\space.\else
  \ifx\@let@token\@sptoken.\else
   .\space
   \fi\fi\fi\fi\fi\fi\fi\fi\fi\fi\fi\fi\fi\fi\fi\fi\fi\fi}
\catcode`\@=12 

\newcommand{\stru}[2]{%
   \relax\ifmmode\hbox{\vrule height#1 depth#2 width0pt}%
   \else\vrule height#1 depth#2 width0pt\fi}

\newcommand{\Ronum}[1]{\uppercase\expandafter{\romannumeral#1}}
\newcommand{\ronum}[1]{\expandafter{\romannumeral#1}}
\DeclareRobustCommand{\LaTeXZ}{%
  \LaTeX\kern-.05em4\kern-.1em
  {\raisebox{-0.2ex}{$\scriptstyle\text{ZEUS}$}}\xspace}

\DeclareMathAlphabet{\mathbf}{OT1}{cmr}{bx}{sl}
\newcommand{\eVdist}{\kern-0.06667em}

\newcommand{\gev}{{\,\text{Ge}\eVdist\text{V\/}}}

\newcommand{\slashfrac}[2]{%
  \raisebox{0.5ex}{\ensuremath #1}\kern-0.12em/\kern-0.08em
  \raisebox{-.8ex}{\ensuremath #2}}

\newcommand{\sqr}[3]{%
    {\vcenter{\hrule height.#3ex\hbox{\vrule width.#2ex height#1ex
     \kern#1ex\vrule width.#3ex}\hrule height.#2ex}}}

\catcode`\@=11 
\newcommand{\parenbar}{\mathpalette\p@renb@r}
\def\p@renb@r#1#2{\vbox{%
  \ifx#1\scriptscriptstyle \dimen@.7em\dimen@ii.2em\else
  \ifx#1\scriptstyle \dimen@.8em\dimen@ii.25em\else
  \dimen@1em\dimen@ii.4em\fi\fi \offinterlineskip
  \ialign{\hfill##\hfill\cr
    \vbox{\hrule width\dimen@ii}\cr
    \noalign{\vskip-.3ex}%
    \hbox to\dimen@{$\mathchar300\hfil\mathchar301$}\cr
    \noalign{\vskip-.3ex}%
    $#1#2$\cr}}}
\catcode`\@=12 

\newcommand{\IP}{{\rm I$\kern-0.01667em$P}\xspace}

\mathchardef\qsm=63
\mathchardef\pls=43
\mathchardef\mns=512
\mathchardef\plm=518
\mathchardef\eql=61
\mathchardef\smallleft=300
\mathchardef\smallright=301
\mathchardef\les=316
\mathchardef\gre=318
\mathchardef\leq=532
\mathchardef\grq=533

\catcode`\@=11 
\newcounter{pict@width}
\newcounter{pict@height}
\newlength{\pict@scale}
\newcommand{\psfigadd}[4]{%
\setcounter{pict@width}{1*\ratio{#2+\pict@scale/2}{\pict@scale}}
\setcounter{pict@height}{1*\ratio{#3+\pict@scale/2}{\pict@scale}}
\setlength{\unitlength}{\pict@scale}
\hbox to #2{\hspace{-\fill}\begin{picture}(\thepict@width,\thepict@height)
\put(0,0){\psfig{figure=#1,width=#2,height=#3,clip=}}
\SetScale{0.283466457}
\SetWidth{1.763889}
{#4}
\end{picture}}
}
\newcounter{pict@widthfst}
\newcounter{pict@widthscd}
\newcounter{pict@widthtot}
\newcommand{\psfigaddtwo}[7]{%
\setcounter{pict@widthfst}{1*\ratio{#2+\pict@scale/2}{\pict@scale}}
\setcounter{pict@widthscd}{1*\ratio{#2+#4+\pict@scale/2}{\pict@scale}}
\setcounter{pict@widthtot}{1*\ratio{#2+#4+#6+\pict@scale/2}{\pict@scale}}
\setcounter{pict@height}{1*\ratio{#3+\pict@scale/2}{\pict@scale}}
\setlength{\unitlength}{\pict@scale}
\hbox{\hspace{-\fill}\begin{picture}(\thepict@widthtot,\thepict@height)
\put(0,0){\psfig{figure=#1,width=#2,height=#3,clip=}}
\put(\thepict@widthscd,0){\psfig{figure=#5,width=#6,height=#3,clip=}}
\SetScale{0.283466457}
\SetWidth{1.763889}
{#7}
\end{picture}}
}
\newcommand{\psfigror}[4]{%
\setcounter{pict@width}{1*\ratio{#2+\pict@scale/2}{\pict@scale}}
\setcounter{pict@height}{1*\ratio{#3+\pict@scale/2}{\pict@scale}}
\setlength{\unitlength}{\pict@scale}
\hbox{\begin{picture}(\thepict@width,\thepict@height)
\put(0,\thepict@height){\psfig{figure=#1,width=#3,height=#2,clip=,angle=270}}
\SetScale{0.283466457}
\SetWidth{1.763889}
{#4}
\end{picture}}
}
\newcommand{\psfigrol}[4]{%
\setcounter{pict@width}{1*\ratio{#2+\pict@scale/2}{\pict@scale}}
\setcounter{pict@height}{1*\ratio{#3+\pict@scale/2}{\pict@scale}}
\setlength{\unitlength}{\pict@scale}
\hbox{\begin{picture}(\thepict@width,\thepict@height)
\put(0,0){\psfig{figure=#1,width=#3,height=#2,clip=,angle=90}}
\SetScale{0.283466457}
\SetWidth{1.763889}
{#4}
\end{picture}}
}
\catcode`\@=12 
\newlength\listtextwidth

\catcode`\@=11 
\newlength{\@tabfninsert}
\newlength{\@tabfnwidth}
\newcommand{\tabfootnote}[2]{%
  \setlength{\@tabfninsert}{0.8em}
  \setlength{\@tabfnwidth}{\textwidth}
  \addtolength{\@tabfnwidth}{-\@tabfninsert}
  \addtolength{\@tabfnwidth}{-0.4em}
  \noindent\makebox[\@tabfninsert][r]{\footnotesize$^{#1}$\hfil}\hfill%
  \parbox[t]{\@tabfnwidth}{\footnotesize #2\hfill}}
\catcode`\@=12 

\newcommand{\leqsim}{\,\raisebox{-0.6ex}{$\buildrel < \over \sim$}\,}

\newcommand{\msbar}{\mbox{$\overline{\rm{MS}}$}\ }
\begin{document}
\prepnum{{DESY-05-050}}\title{An NLO QCD analysis of inclusive cross-section and jet-production data from the ZEUS experiment
}
                    
\author{ZEUS Collaboration}

\date{Mar 2005}

\abstract{The ZEUS inclusive differential cross-section data from HERA, 
   for charged and neutral
   current processes taken with $e^+$ and $e^-$ beams, together with 
   differential cross-section data on inclusive
   jet production in $e^+ p$ scattering and dijet production in $\gamma p$
   scattering, have been used in a new NLO QCD analysis to extract 
   the parton distribution functions of the proton.
   The input of jet-production data constrains the gluon and  
   allows an accurate extraction of $\alpha_s(M_Z)$ at NLO;
\[  \alpha_s(M_Z) = 0.1183 \pm 0.0028({\rm exp.}) \pm 0.0008({\rm model}) .
\] 
   An additional uncertainty from the choice of scales is 
   estimated as $\pm 0.005$. This is the first extraction of $\alpha_s(M_Z)$
   from HERA data alone.

}

\makezeustitle

\def\3{\ss}                                                                                        
\pagenumbering{Roman}                                                                              
                                                   %
\begin{center}                                                                                     
{                      \Large  The ZEUS Collaboration              }                               
\end{center}                                                                                       
  S.~Chekanov,                                                                                     
  M.~Derrick,                                                                                      
  S.~Magill,                                                                                       
  S.~Miglioranzi$^{   1}$,                                                                         
  B.~Musgrave,                                                                                     
  \mbox{J.~Repond},                                                                                
  R.~Yoshida\\                                                                                     
 {\it Argonne National Laboratory, Argonne, Illinois 60439-4815}, USA~$^{n}$                       
\par \filbreak                                                                                     
  M.C.K.~Mattingly \\                                                                              
 {\it Andrews University, Berrien Springs, Michigan 49104-0380}, USA                               
\par \filbreak                                                                                     
  N.~Pavel, A.G.~Yag\"ues Molina \\                                                                
  {\it Institut f\"ur Physik der Humboldt-Universit\"at zu Berlin,                                 
           Berlin, Germany}
% ...
\newpage                                                                                           
$^{\    1}$ also affiliated with University College London, UK \\                                  
% ...
                                                           %
                                                           %
\newpage   
                                                           %
                                                           %
\begin{tabular}[h]{rp{14cm}}                                                                       
$^{a}$ &  supported by the Natural Sciences and Engineering Research Council of Canada (NSERC) \\  
$^{b}$ &  supported by the German Federal Ministry for Education and Research (BMBF), under        
          contract numbers HZ1GUA 2, HZ1GUB 0, HZ1PDA 5, HZ1VFA 5\\                                
$^{c}$ &  supported in part by the MINERVA Gesellschaft f\"ur Forschung GmbH, the Israel Science   
          Foundation (grant no. 293/02-11.2), the U.S.-Israel Binational Science Foundation and    
          the Benozyio Center for High Energy Physics\\                                            
$^{d}$ &  supported by the German-Israeli Foundation and the Israel Science Foundation\\           
$^{e}$ &  supported by the Italian National Institute for Nuclear Physics (INFN) \\                
$^{f}$ &  supported by the Japanese Ministry of Education, Culture, Sports, Science and Technology 
          (MEXT) and its grants for Scientific Research\\                                          
$^{g}$ &  supported by the Korean Ministry of Education and Korea Science and Engineering          
          Foundation\\                                                                             
$^{h}$ &  supported by the Netherlands Foundation for Research on Matter (FOM)\\                   
$^{i}$ &  supported by the Polish State Committee for Scientific Research, grant no.               
          620/E-77/SPB/DESY/P-03/DZ 117/2003-2005 and grant no. 1P03B07427/2004-2006\\             
$^{j}$ &  partially supported by the German Federal Ministry for Education and Research (BMBF)\\   
$^{k}$ &  supported by RF Presidential grant N 1685.2003.2 for the leading scientific schools and  
          by the Russian Ministry of Education and Science through its grant for Scientific        
          Research on High Energy Physics\\                                                        
$^{l}$ &  supported by the Spanish Ministry of Education and Science through funds provided by     
          CICYT\\                                                                                  
$^{m}$ &  supported by the Particle Physics and Astronomy Research Council, UK\\                   
$^{n}$ &  supported by the US Department of Energy\\                                               
$^{o}$ &  supported by the US National Science Foundation\\                                        
$^{p}$ &  supported by the Polish Ministry of Scientific Research and Information Technology,      
          grant no. 112/E-356/SPUB/DESY/P-03/DZ 116/2003-2005 and 1 P03B 065 27\\                  
$^{q}$ &  supported by FNRS and its associated funds (IISN and FRIA) and by an Inter-University    
          Attraction Poles Programme subsidised by the Belgian Federal Science Policy Office\\     
\end{tabular}                                                                                      
                                                           %

\newpage
\section{Introduction} 

Since the advent of HERA, considerable progress has been made in the 
determination of the parton distribution functions (PDFs) of the proton.
Precise knowledge of the PDFs, and of the strong coupling constant, 
$\alpha_s(M_Z)$, is crucial for an 
understanding of proton structure. Moreover, it is required for any 
calculation 
of cross sections at hadron colliders both for Standard Model physics and for
the discovery of physics beyond the Standard Model.

The PDFs are usually determined in global 
fits~\cite{epj:c23:73,epj:c28:455,jhep:0207:012} made within the conventional 
DGLAP formalism~\cite{sovjnp:15:438,sovjnp:20:94,np:b126:298,jetp:46:641} 
at next-to-leading order (NLO). Such fits 
use data from many different experiments, with the 
inclusive cross-section data from deep inelastic scattering (DIS) experiments 
providing the major source of information. The wide kinematic range covered by
the HERA DIS data~\cite{pr:d67:012007,epj:c21:33,epj:c30:1}, as well as their 
precision,
has allowed the determination of PDFs across a broad range of phase space 
spanned by the fractional proton momentum carried by the struck quark, 
Bjorken $x$, and the 
negative squared four-momentum transfer between the lepton and nucleon, $Q^2$.
The high-statistics HERA neutral current 
$e^+p$ data determine the low-$x$ sea and 
gluon distributions, whereas the fixed-target data, taken at lower 
centre-of-mass energy, determine 
the valence distributions and the higher-$x$ sea distributions. 

The gluon PDF contributes only indirectly to the 
inclusive DIS cross sections. However it  makes a direct contribution to
jet cross sections through boson-gluon and quark-gluon 
scattering. Tevatron high-$E_T$ jet data~\cite{prl:82:2451,pr:d64:032001} 
have been used to constrain the gluon in the fits of 
MRST~\cite{epj:c23:73,epj:c28:455} and CTEQ~\cite{jhep:0207:012}. However, 
these data suffer from very large correlated systematic uncertainties 
from a variety of sources. For example, the total systematic uncertainty of 
CDF data is $\sim 60\%$ over its full $E_T$ range.
In the present paper, ZEUS neutral current $e^+p$ DIS inclusive 
jet cross sections~\cite{pl:b547:164} and direct 
photoproduction dijet cross sections~\cite{epj:c23:615} have been used to 
constrain the gluon. These cross sections
have only $\sim 5\%$ total systematic uncertainty, 
mainly due to the absolute energy-scale uncertainty of the jets. 

These jet data were used, together with 
ZEUS data on neutral and charged current (NC and CC) 
$e^+p$ and $e^-p$ DIS inclusive  
cross sections~\cite{epj:c21:443,epj:c12:411,epj:c28:175,pl:b539:197,
pr:d70:052001,epj:c32:16}, as inputs to an 
NLO QCD DGLAP
analysis in order to determine the PDFs. 
This fit is called the ZEUS-JETS fit.

In the ZEUS-JETS fit, 
the lower $Q^2$ NC inclusive cross-section data determine 
the low-$x$ sea and
gluon distributions\footnote{The HERA kinematics is such that the lower-$Q^2$ 
data are also at low $x$.} and  the high $Q^2$ NC and CC inclusive 
cross sections determine the valence distributions. 
The use of ZEUS data alone eliminates the uncertainty from heavy-target 
corrections required in global analyses in which
the $\nu$\emph{Fe}  and $\mu$\emph{D}  
fixed-target data, together with isospin-symmetry 
constraints between $u$ and $d$ in the proton and neutron, have been used for 
determining the valence distributions.
The jet cross-section data constrain the mid- to high-$x$ 
($x \approx 0.01 - 0.5$) gluon PDF. The
predictions for the jet cross sections are calculated to NLO in QCD and 
are used in the fit rigorously, rather than 
approximately as in previous fits~\cite{epj:c23:73,epj:c28:455,jhep:0207:012}.
The quality of the fit 
establishes that NLO QCD is able simultaneously to describe both 
inclusive cross sections and jet cross sections, thereby
providing a compelling demonstration of QCD factorisation.

 The value of $\alpha_s(M_Z)$
is fixed in most PDF fits; for the ZEUS-JETS fit, the value 
$\alpha_s(M_Z) =  0.118$~\cite{pl:b592:1} is used. 
A simultaneous fit for $\alpha_s(M_Z)$ and the PDF parameters, called the 
ZEUS-JETS-$\alpha_s$ fit, has also been made. This fit accounts for the 
correlation between $\alpha_s(M_Z)$ and the gluon shape. The addition of the 
jet production data provides enough constraints to give an accurate 
determination of $\alpha_s(M_Z)$ despite this correlation. 

The PDFs are presented with full accounting for uncertainties from correlated 
systematic errors.
Performing an analysis within a single experiment has considerable advantages
in this respect since global fits have found significant tensions between 
different data sets~\cite{epj:c23:73}.
In the present analysis, the contribution to the PDF 
uncertainties from correlated experimental uncertainties and normalisation
uncertainties is significantly reduced in comparison to the previous ZEUS-S 
global fit analysis~\cite{pr:d67:012007}, which used data from many 
different DIS experiments.

This paper is organised as follows. In Section~\ref{sec:theo}, the theoretical 
background is reviewed briefly and in Section~\ref{sec:anal}, 
the method of analysis is outlined,
 with particular emphasis on the new features needed to include the jet 
cross sections in the fit. 
In Section~\ref{sec:res}, the ZEUS-JETS fit is compared 
to data and the extracted parton distributions and their experimental 
uncertaintes are presented. Model uncertainties are discussed and a 
comparison is made to the Tevatron jet data. In Section~\ref{sec:alphas}, 
the analysis is extended to the evaluation of $\alpha_s(M_Z)$ in the
ZEUS-JETS-$\alpha_s$ fit and the 
uncertainties on $\alpha_s(M_Z)$ from theoretical sources are discussed.
Section~\ref{sec:summ} gives a summary and conclusions.

\section{Theoretical Background}
\label{sec:theo}

The kinematics of deep inelastic lepton-nucleon scattering are described in 
terms of the variables $Q^2$, 
Bjorken $x$ and $y$, the fractional energy transfer between the
lepton and hadron systems.
The differential cross sections for the NC DIS process are given in terms of 
structure functions by
\[
\frac {d^2\sigma^{\rm NC}(e^{\pm}p) } {dxdQ^2} =  \frac {2\pi\alpha^2} {x Q^4}
\left[Y_+\,F_2(x,Q^2) - y^2 \,F_L(x,Q^2)
\mp Y_-\, xF_3(x,Q^2) \right],
\]
where $\displaystyle Y_\pm=1\pm(1-y)^2$. 
The structure functions $F_2$ and $xF_3$ are 
directly related to quark distributions, and their
$Q^2$ dependence, or scaling violation, 
is predicted by perturbative QCD. 
At $Q^2 \leqsim 1000$~GeV$^2$, the
charged lepton-hadron cross section is dominated by photon exchange and the 
structure function $F_2$. For $x \leqsim 10^{-2}$, $F_2$  
is sea-quark dominated and its $Q^2$ dependence is driven by
the gluon contribution, such that HERA data provide 
crucial information on both quark and gluon distributions.
The longitudinal structure function $F_L$ is only important at high $y$ and
is calculated, in perturbative QCD, 
from the quark and gluon distributions~\cite{pl:b76:89}.
At high $Q^2$, the structure function $xF_3$ becomes increasingly important; 
it  provides information on valence quark distributions. The CC interactions 
are sensitive to the flavour of the valence distributions 
at high $x$  since their (LO) cross sections are given by
\[
\frac {d^2\sigma^{\rm CC}(e^+ p) } {dxdQ^2} = \frac {G_F^2 M_W^4} 
{2\pi x (Q^2 +M_W^2)^2} x\left[(\bar{u}+\bar{c}) + (1 - y)^2 (d + s) \right],
\]
\[
\frac {d^2\sigma^{\rm CC}(e^- p) } {dxdQ^2} = \frac {G_F^2 M_W^4} 
{2\pi x (Q^2 +M_W^2)^2} x\left[(u + c) + (1 - y)^2 (\bar{d} + \bar{s}) \right],
\]
where the parton distributions $u$, $d$, $s$, $c$ are functions of $x$ and 
$Q^2$.
Thus the $e^- p$ CC cross section gives information on the $u$ valence quark
at high $x$, whereas the $e^+ p$ CC cross section gives information on the
$d$ valence quark at high $x$. This is particularly important since this 
process  is a direct probe of the $d$ valence quark on a proton target at 
high $Q^2$. Determinations of the 
$d$ valence distribution have previously been dominated by low $Q^2$ data 
using isoscalar iron or deuterium targets. Such determinations
are subject to uncertainties from higher-twist contributions, 
heavy-target and binding corrections and isospin-symmetry assumptions.

The inclusive cross-section data depend directly on the quark distributions,
but the gluon distribution affects these cross sections indirectly through
the scaling violations. 
Perturbative QCD predicts the rate at which the quark distributions evolve 
with the scale $Q$ through the DGLAP equation
\begin{equation}
\frac {d q_i(x,Q^2)} {d \ln Q^2} = \frac {\alpha_s(Q^2)} {2\pi} 
\int_x^1 \frac {dy}{y} \left[\sum_j
q_j(y,Q^2) P_{q_iq_j}\left(\frac{x}{y}\right) + g(y,Q^2) P_{q_ig} \left(\frac{x}{y}\right) \right],
\end{equation}
where the `splitting function' $P_{ij}(z)$ represents the probability of 
a parton
(either quark or gluon) $j$ emitting a quark $i$ with momentum fraction $z$
of that of the parent parton.
Thus the gluon distribution can be obtained indirectly from 
the scaling violations of the quark distributions. 
The parameters that describe the gluon shape and the value of the strong 
coupling constant, $\alpha_s(M_Z)$, are 
correlated through the DGLAP equations. 

The QCD processes that give rise to scaling violations in the 
inclusive cross sections, namely the QCD-Compton (QCDC) and boson-gluon-fusion
(BGF) processes, are observed as events with distinct jets in the final 
state provided that the energy and momentum transfer are large enough.  The 
cross section for QCDC scattering depends on $\alpha_s(M_Z)$ and the quark 
PDFs. For HERA kinematics, 
this process dominates the jet cross section at large scales, 
where the quark densities are well known from the inclusive cross-section 
data, so that the value of $\alpha_s(M_Z)$ may be extracted without strong
correlation to the shape of the gluon PDF. The cross section for the
BGF process depends on $\alpha_s(M_Z)$ and the gluon PDF so that measurements 
of jet cross sections also provide a direct determination of the gluon density.

\section{Analysis Method}
\label{sec:anal}

The present analysis was performed within the Standard Model conventional
paradigm of leading-twist NLO QCD. The QCD predictions for the PDFs 
were obtained by solving the DGLAP evolution equations at NLO.  
These equations yield the PDFs
at all values of $Q^2$ provided they
are parameterised as functions of $x$ at some input scale $Q_0$. 
The programme {\sc Qcdnum}~\cite{upub:botje:qcdnum1612} 
was used to perform the evolution.

The applicability of the leading-twist, NLO DGLAP formalism to HERA data 
was investigated in the previous ZEUS 
analysis~\cite{pr:d67:012007}, and suitable data cuts 
were defined. All the present data lie above these cuts.
The data sets fitted in this analysis and their kinematic coverage 
are presented in Table~\ref{tab:chisq}. In total there are 
577 data points from a total luminosity of  112 pb$^{-1}$ from the 
HERA-I (1992-2000) running period.

Full account has been taken of correlated experimental 
systematic uncertainties using the Offset method, 
described in the previous ZEUS-S PDF analysis~\cite{pr:d67:012007}\footnote{ 
Different treatments of experimental uncertainties in PDF analyses are 
discussed extensively elsewhere~\cite{jp:g28:779,jp:g28:2609,jp:g28:2717}. 
The Offset method gives conservative PDF uncertainty estimates.}.
There are 22 independent sources of correlated systematic uncertainty 
and 4 independent normalisations for the data sets in the present 
analysis. The number of correlated systematic uncertainties for each data set,
their normalisations and the correlations between the data sets 
are detailed in Table~\ref{tab:chisq}.

\subsection{Inclusive cross-section data}

The inclusive cross-section data used in the fits were reduced double 
differential cross-sections in $x$ 
and $Q^2$ from: NC $e^+ p$ scattering~\cite{epj:c21:443,pr:d70:052001}; 
NC $e^- p$ scattering~\cite{epj:c28:175}; CC $e^+ p$ 
scattering~\cite{epj:c12:411,epj:c32:16}; and CC $e^- p$ 
scattering~\cite{pl:b539:197}.

The NLO QCD predictions for the structure functions, which enter into the 
expressions for the cross sections, were obtained by convoluting the PDFs
with the QCD coefficient functions appropriate to the process. 
It is necessary to specify the scheme and scale choice for the calculations.
The renormalisation and factorisation scales for the inclusive DIS processes 
were chosen to be $Q$. The DGLAP equations were solved in the \msbar\ scheme.
For heavy-quark production, the general-mass variable 
flavour-number scheme of Thorne and Roberts (TRVFN)~\cite{pr:d57:6871} was 
used in order to interpolate correctly between threshold behaviour and  
high-scale behaviour for heavy quarks, as discussed in the ZEUS-S 
analysis~\cite{pr:d67:012007}. The values of the heavy quark masses used were
$m_c=1.35$~GeV and $m_b=4.3$~GeV. Variation of these values in the ranges
$1.2 < m_c < 1.5$~GeV and $4.0 < m_b < 4.6$~GeV 
produced changes in the PDF parameters 
that are negligible in comparison to the experimental uncertainties.

\subsection{Jet data}
\label{sec:jets}

The jet data used in the fits were: 
DIS inclusive jet differential cross sections as a function of the 
transverse energy  in the 
Breit frame, $E_T^B$, for different $Q^2$ bins~\cite{pl:b547:164};
photoproduction dijet 
cross sections as a function of the transverse energy of the most energetic 
jet, $E_T^{\rm jet1}$, in the laboratory frame, 
for different jet-pseudorapidity ranges~\cite{epj:c23:615}.
The systematic uncertainty from the absolute jet energy scale was 
fully correlated between these two sets of data.

The cross-section predictions for photoproduced jets are sensitive to the 
choice of the input photon PDFs. The AFG photon 
PDF~\cite{zfp:c64:621} has been 
used in the fits. In order to minimise sensitivity to this choice, the 
analysis has been restricted to use only the `direct' photoproduction 
cross sections. These are defined by the cut $x^{\rm obs}_\gamma > 0.75$, 
where $x^{\rm obs}_\gamma$ is a measure of the fraction of the photon's 
momentum that enters into 
the hard scatter~\cite{epj:c23:615,pl:b348:665,pl:b322:287}.

The programme of Frixione and Ridolfi~\cite{np:b507:315} was used to compute 
NLO QCD cross sections for photoproduced dijets and 
{\sc Disent}~\cite{np:b510:503}
was used to compute NLO QCD cross sections for jet production in DIS. 
These programmes treat the heavy quarks in a massless scheme. However all the
jet data are at scales sufficiently high that the TRVFN scheme
 and the zero-mass variable flavour number scheme (ZMVFN) are equivalent.
The calculation of the NLO jet cross sections 
was too slow to be used iteratively in the fit. 
Thus, they were used to compute LO and NLO weights, $\tilde{\sigma}$, 
which are independent of $\alpha_s$ and the 
PDFs, and are obtained by integrating the corresponding partonic hard cross 
sections\footnote{For the dijet photoproduction cross sections,
the weights also included the convolution with the photon PDFs.} in
bins of $\xi$ (the proton momentum fraction carried by
the incoming parton), $\mu_F$ (the factorisation scale) and,
for the case $\mu_F \ne \mu_R$, $\mu_R$ (the renormalisation scale).
The NLO QCD cross sections, for each measured bin, were then 
obtained by folding these weights with the PDFs and $\alpha_s$ according to the
formula
\begin{equation}
         \sigma = \sum_n \sum_a \sum_{i,j,k} 
                  f_a({\langle \xi \rangle}_i , {\langle \mu_F \rangle}_j)  
                  \cdot \alpha_s^n({\langle \mu_R \rangle}_k) 
                  \cdot \tilde{\sigma}^{(n)}_{a,\{i,j,k\}} ~,
\end{equation}
where the three sums run over the order $n$ in $\alpha_s$, the flavour $a$ of 
the incoming parton, and the indices ($i,j,k$) of the $\xi$, $\mu_F$ and 
$\mu_R$ bins, respectively. 
The PDF, $f_a$, and $\alpha_s$ were evaluated at the mean values 
${\langle \xi \rangle}$, ${\langle \mu_F \rangle}$ and
${\langle \mu_R \rangle}$ of the variables $\xi$,~$\mu_F$ and $\mu_R$ in each 
($i,j,k$) bin. The factorisation scale was chosen as
$\mu_F= Q$ for the DIS jets, and the renormalisation scale was chosen as 
$\mu_R=E_T^B$ (with
$\mu_R=Q$ as a cross-check). For the photoproduced dijets, the standard 
scale choices were $\mu_R= \mu_F = E_T/2$  (where $E_T$  is the 
summed transverse momenta of final-state partons). 
This procedure reproduces the NLO predictions
 to better than $0.5\%$.

The predictions were multiplied by 
hadronisation corrections before they were used to fit the data. 
These were determined by using  Monte Carlo (MC) programmes, 
which model parton hadronisation to estimate the ratio of 
the hadron- to parton-level cross sections for each bin. For the DIS jet data,
an average of the values obtained using the {\sc Ariadne}, 
{\sc Lepto} and {\sc Herwig} 
MC programmes was taken~\cite{pl:b547:164}. For the photoproduction data, an 
average of the values obtained 
from the {\sc Herwig} and {\sc Pythia} MC
programmes was taken~\cite{epj:c23:615}. 
The hadronisation corrections are generally within a few 
percent of unity~\cite{pl:b547:164,epj:c23:615}.
The predictions for DIS jet production were also corrected for 
$Z^0$ contributions.

\subsection{Parameterisation of PDFs}

The PDFs for $u$ valence,  $d$ valence, 
total sea, gluon and the difference between the $d$ and $u$
contributions to the sea, are each parameterised, at $Q^2_0 = 7$~GeV$^2$, 
by the form 
\[
  xf(x) = p_1 x^{p_2} (1-x)^{p_3}( 1 + p_4 x).
\]
It was checked that no significant improvement 
in $\chi^2$ results from the use of 
more complex polynomial forms or from variation of  
the value of $Q^2_0$. The following constraints were imposed on the 
parameters $p_i$:
\begin{itemize}
\item the normalisation parameters $p_1$, for the $d$ 
and $u$ valence and for the gluon, were constrained by imposing the 
number sum-rules and momentum sum-rule, respectively; 
\item the $p_2$ parameters, which constrain the low-$x$ behaviour of 
the valence distributions, were set equal for $u$ and $d$, 
since there is insufficient information to constrain any difference; 
\item there is 
also no information on the flavour structure of the light-quark sea in a fit 
to ZEUS data alone. Thus, the normalisation of the $\bar{d}-\bar{u}$ 
distribution was fixed to be consistent with the measured violation of 
the Gottfried sum-rule~\cite{pl:b295:159,prl:66:2712} and its shape was 
fixed to be consistent with 
the Drell-Yan data~\cite{pr:d64:052002};
\item a suppression of the strange sea by a factor of two at $Q^2_0$ 
was imposed in accordance with neutrino induced dimuon data from 
CCFR-NuTeV~\cite{zfp:c65:189,prl:88:091802}.
\end{itemize} 
The fit is not sensitive to reasonable variations of these assumptions, 
indicating that it is only possible to extract a 
flavour-averaged sea distribution from these ZEUS data.

The ZEUS inclusive cross-section data are statistics limited at large $x$, 
where the sea 
and the gluon distributions are small. This leads to sizeable uncertainties  
in  the mid- to high-$x$ sea and gluon shapes if a fit is made to 
inclusive cross-section data alone.
The  ZEUS jet data constrain the gluon distribution 
in this kinematic region. 
Two different strategies were used to constrain the sea distribution: 
firstly, a simple parameterisation setting $p_4=0$ was used;  
secondly, the $p_4$ parameter was freed but the $p_3$ parameter was 
fixed to the value obtained in the ZEUS-S 
global fit~\cite{pr:d67:012007}. In the latter case, model uncertainties
on the high-$x$ sea include the effect of changing this fixed value of $p_3$ 
within the limits of its uncertainty as determined in the global fit.        
There is very little difference in the shapes and uncertainties 
of the sea  PDF as  determined in these two strategies once 
this model uncertainty on $p_3$ is taken into account. 
Distributions are presented for the former choice because of its simplicity.
Finally, there are 11 free parameters describing the input PDF 
distributions, which are listed in Table~\ref{tab:param}.

\section{Results}

\label{sec:res}

\begin{figure}[ht]
\vspace{-2.0cm} 
\centerline{{\includegraphics[width=14cm]{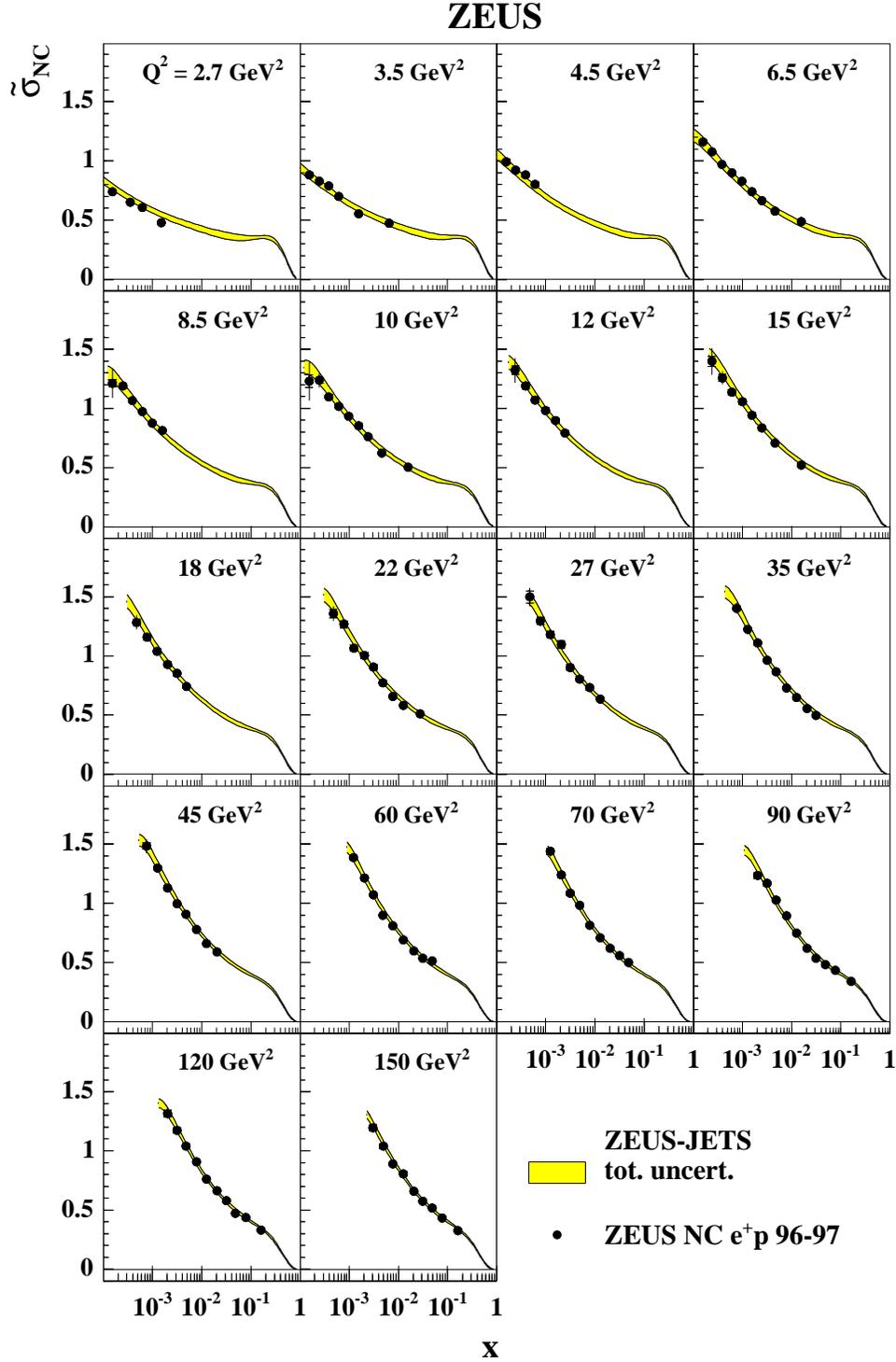}}}
\caption {ZEUS-JETS fit compared to ZEUS low-$Q^2$ $e^+ p$ NC reduced 
cross sections, $\tilde{\sigma}_{\rm NC}$.
}
\label{fig:lowQ2NC}
\end{figure}

\begin{figure}[ht] 
\vspace{-2.0cm} 
\centerline{{\includegraphics[width=14cm]{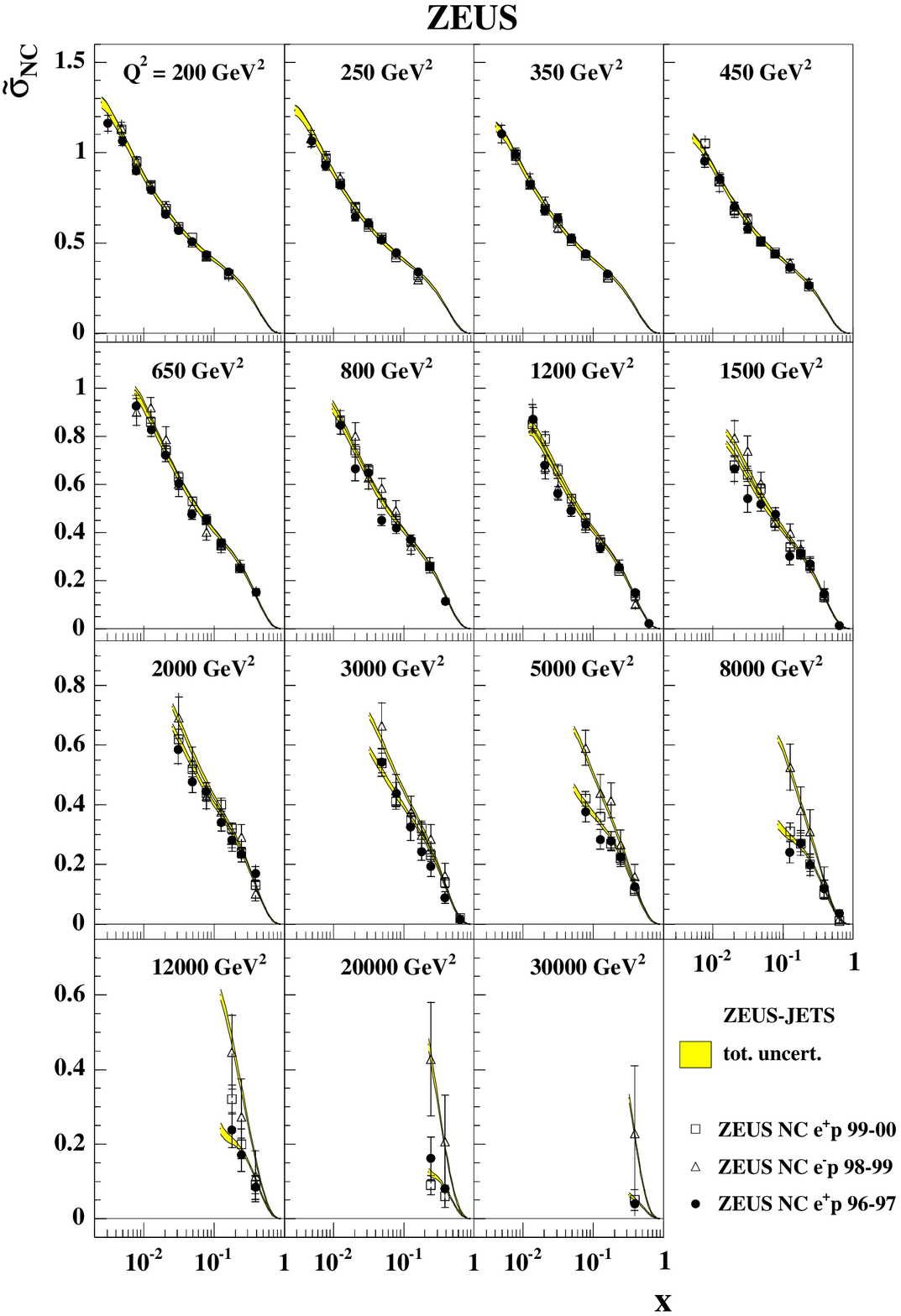}}}
\caption {ZEUS-JETS fit compared to ZEUS high-$Q^2$ NC $e^+ p$ and $e^- p$ 
reduced cross sections, $\tilde{\sigma}_{\rm NC}$.
}
\label{fig:highQ2NC}
\end{figure}

\begin{figure}[ht] 
\vspace{-2.0cm} 
\centerline{{\includegraphics[width=14cm]{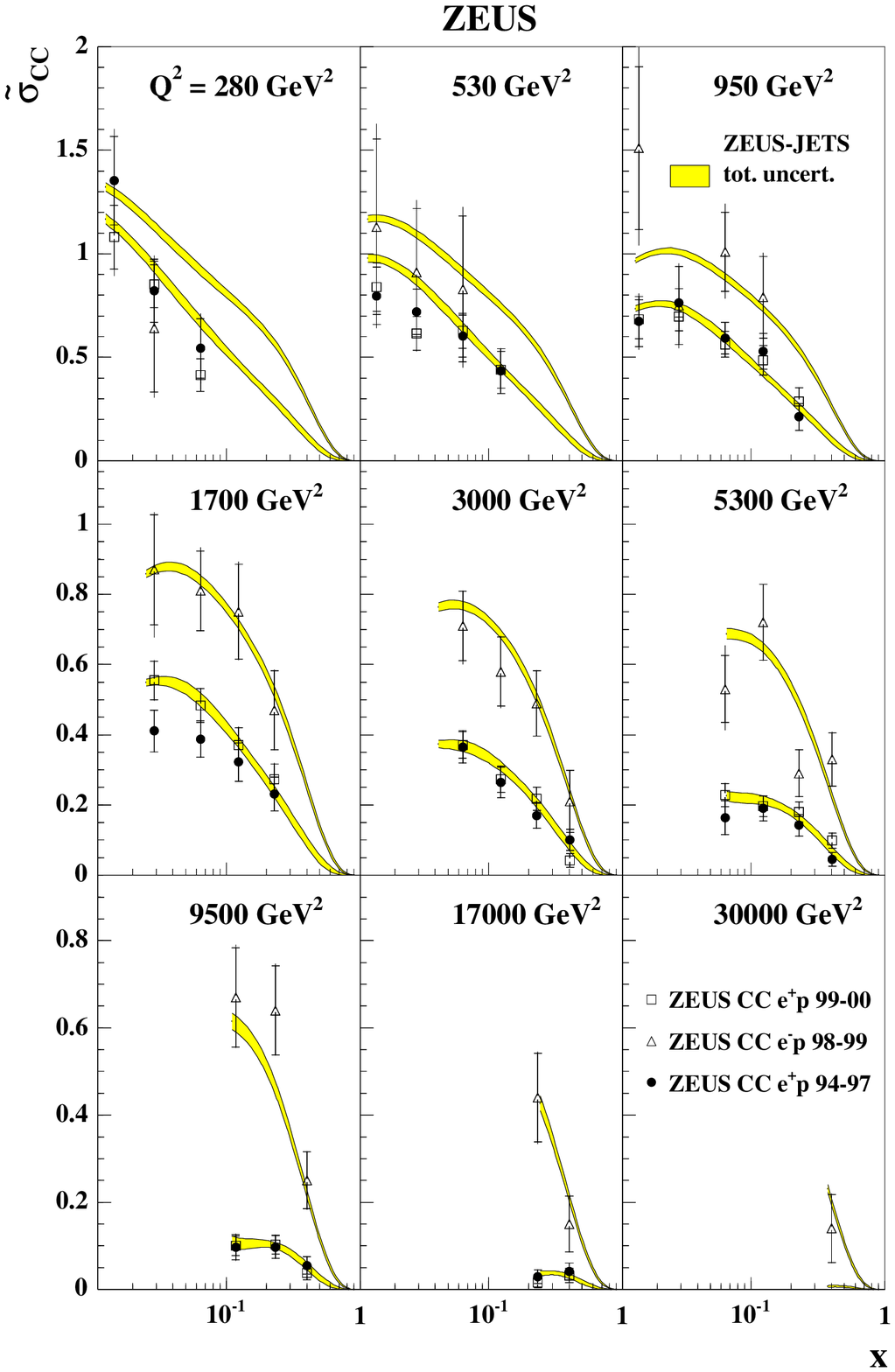}}}
\caption {ZEUS-JETS fit compared to ZEUS high-$Q^2$ CC $e^+ p$ and $e^- p$
reduced cross sections, $\tilde{\sigma}_{\rm CC}$.
}
\label{fig:highQ2CC}
\end{figure}

\begin{figure}[ht] 
\centerline{{\includegraphics[width=12cm]{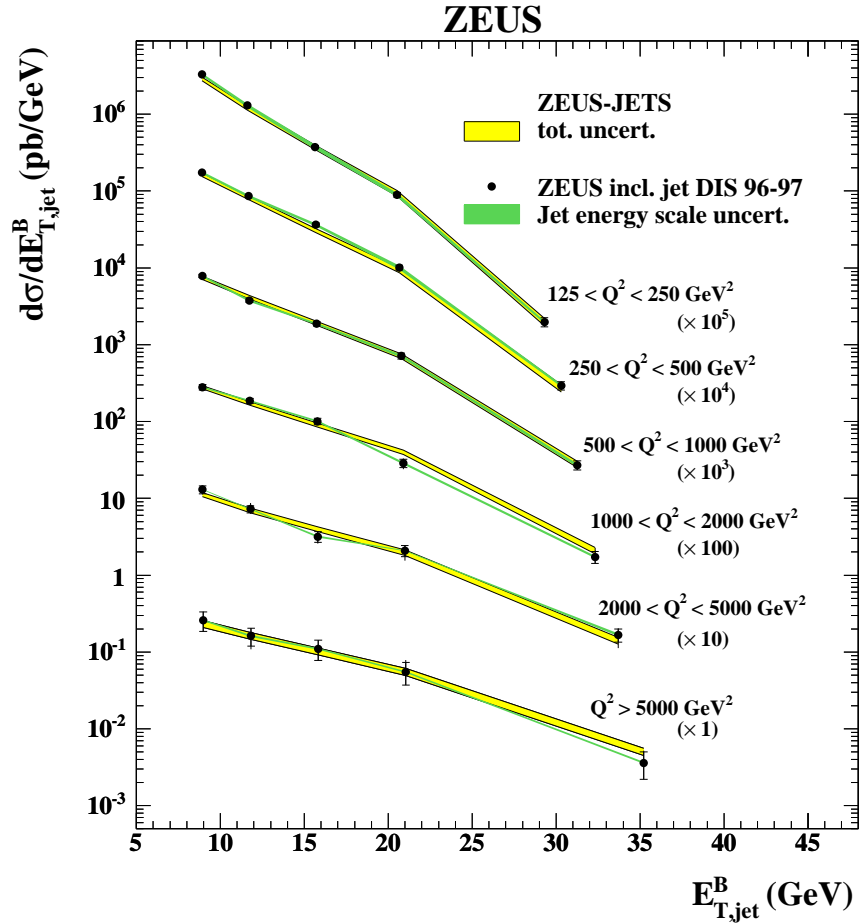}}}
\caption {ZEUS-JETS fit compared to ZEUS DIS jet data. Each cross section 
has been multiplied by the scale factor in brackets to aid visibility.
}
\label{fig:disjets}
\end{figure}

\begin{figure}[ht] 

\centerline{{\includegraphics[width=12cm]{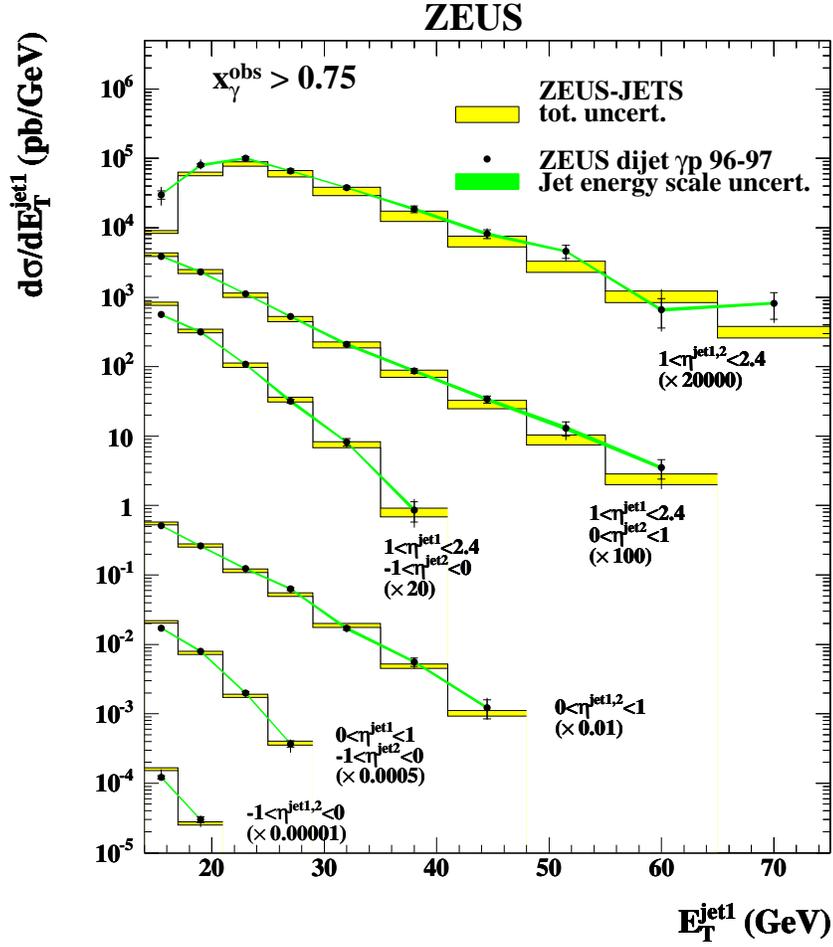}}}
\caption {ZEUS-JETS fit compared to photoproduced dijet data.
Each cross section 
has been multiplied by the scale factor in brackets to aid visibility.
}
\label{fig:phojets}
\end{figure}

The ZEUS-JETS fit and the NC and CC reduced cross-section 
data are shown in 
Figs.~\ref{fig:lowQ2NC},~\ref{fig:highQ2NC} and ~\ref{fig:highQ2CC}. 
The fit and  the jet cross-section data are 
illustrated in Figs.~\ref{fig:disjets} and ~\ref{fig:phojets}. 
A good description of the   
data is obtained over many orders of magnitude in scale.
A measure of the goodness of fit for the Offset method is obtained by 
re-evaluating the  $\chi^2$ by adding the statistical, uncorrelated
and correlated systematic uncertainties 
in quadrature~\cite{jp:g28:779}. The total $\chi^2$ obtained is 470 for 577 
data points.  The extracted PDF parameters  
and their experimental uncertainties are given in Table~\ref{tab:param}.

\begin{figure}[Htp] 
\vspace*{13pt}
\centerline{
\psfig{figure=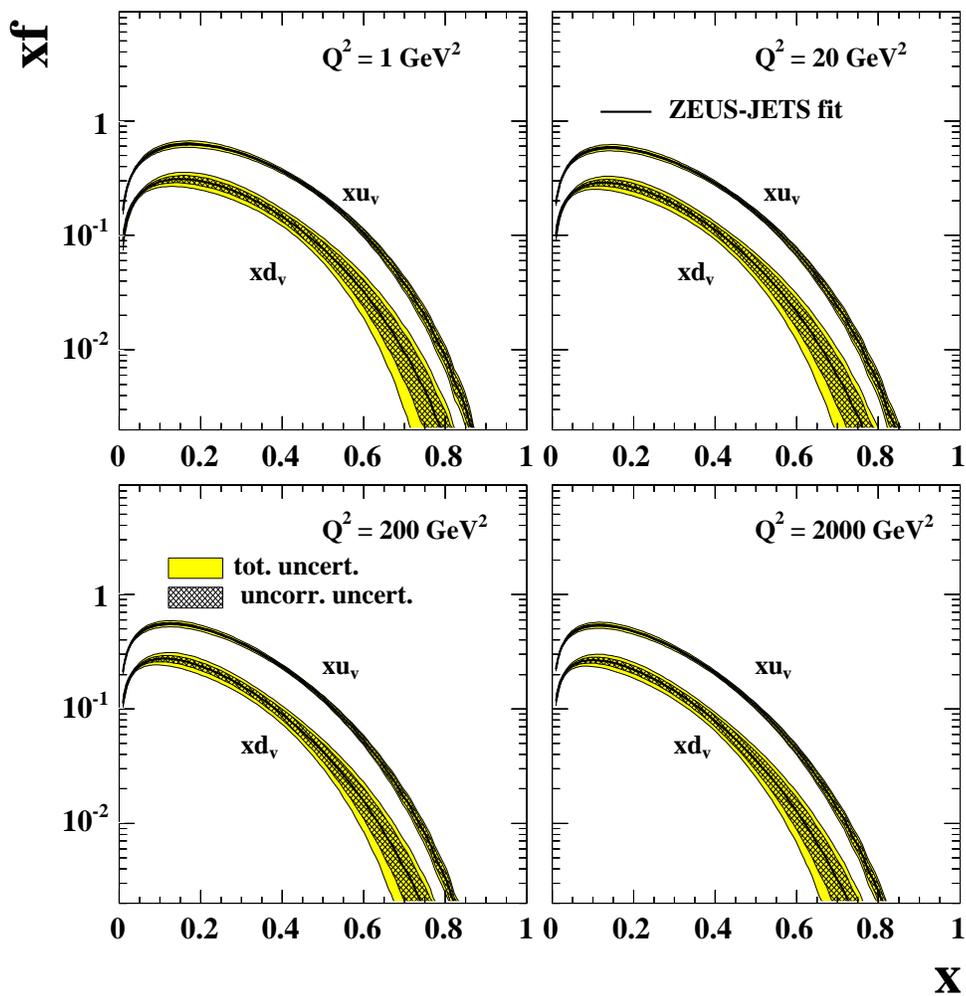,width=0.8\textwidth}
}
\caption {Valence PDFs extracted from the ZEUS-JETS fit. 
The inner cross-hatched 
error bands show the statistical and 
uncorrelated systematic uncertainty, the grey error bands show the 
total uncertainty including experimental correlated systematic uncertainties,
normalisations and model uncertainty.
}
\label{fig:PDFSAvalence}
\end{figure}

The valence distributions for the ZEUS-JETS fit 
are shown in Fig.~\ref{fig:PDFSAvalence}. Although the high-$x$ valence 
distributions are not as well constrained as they are in global fits 
which include fixed-target data,
they are competitive, 
particularly for the less well-known $d$ valence distribution. 
Furthermore, they are free from uncertainties due to heavy-target corrections,
higher-twist effects and isospin-symmetry assumptions.

\begin{figure}[Hbp] 
\vspace*{13pt}
\centerline{
\psfig{figure=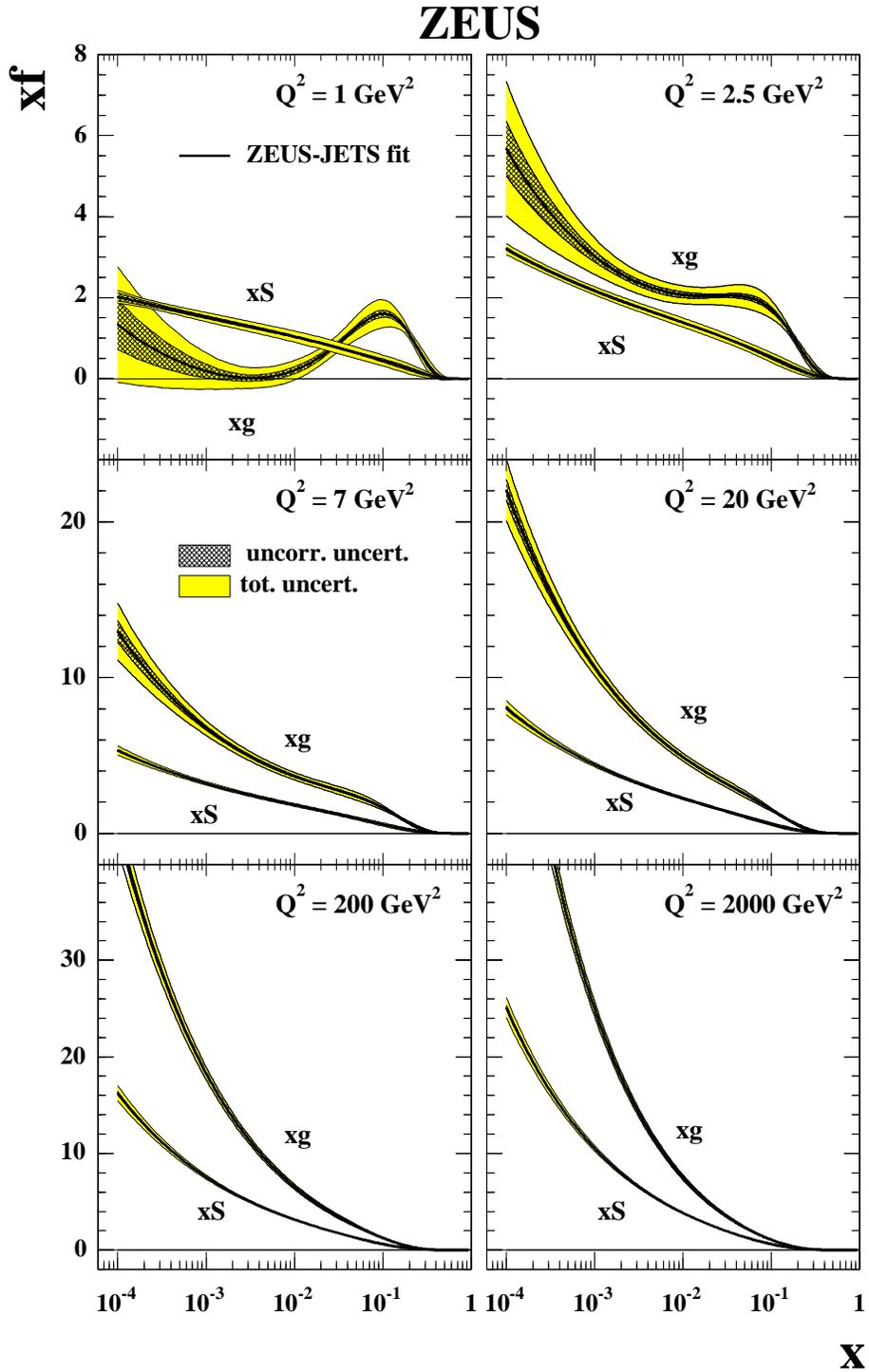,width=0.8\textwidth}}
\caption{
 Gluon and sea PDFs extracted from the ZEUS-JETS fit. 
The uncorrelated and total error bands are as in 
the caption to Fig.~\ref{fig:PDFSAvalence}. 
}
\label{fig:PDFSAglusea}
\end{figure}

The gluon and sea distributions for the ZEUS-JETS fit are shown together in 
Fig.~\ref{fig:PDFSAglusea}. Whereas the sea distribution rises at low $x$ for 
all $Q^2$, the gluon distribution flattens for $Q^2 \sim 2.5$~GeV$^2$ and 
becomes valence-like for lower $Q^2$. 
The gluon and sea distributions are as well determined as the corresponding
distributions of the global 
fits~\cite{pr:d67:012007,epj:c23:73,epj:c28:455,jhep:0207:012} at low $x$ 
since the HERA inclusive NC data 
determine these distributions for all the fits. 
At high $x$, the uncertainties of the sea are constrained to 
be similar to those of the ZEUS-S global fit by the choice of parameterisation,
whereas the uncertainties of the gluon have been reduced by the addition 
of the ZEUS jet data. 

\begin{figure}[tbp] 
\vspace*{13pt}
\centerline{
\psfig{figure=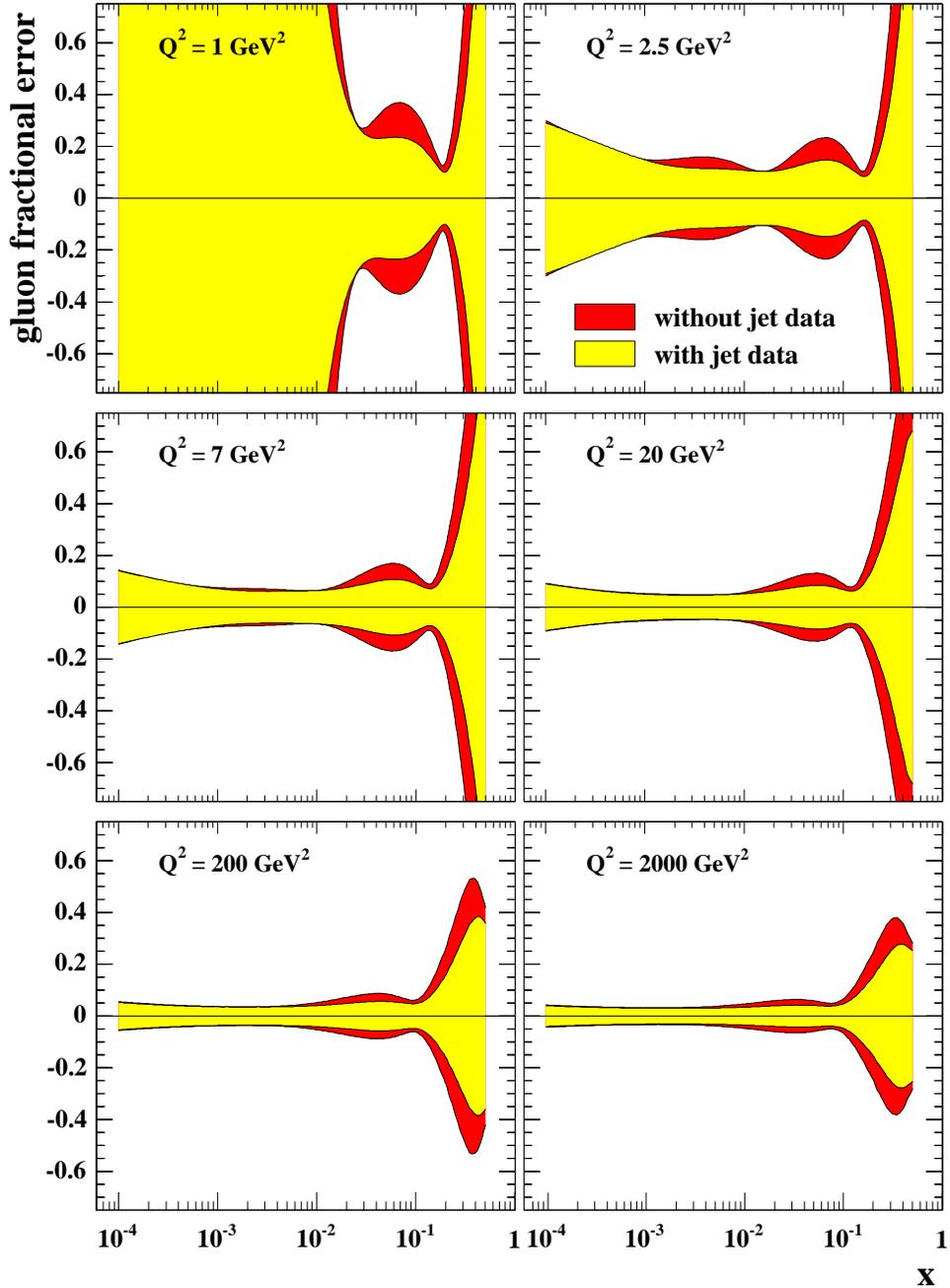,width=0.8\textwidth}}
\caption{
The total experimental uncertainty  on the gluon PDF for 
the ZEUS-JETS fit (central error bands) compared to the total experimental
uncertainty on the gluon PDF for
a fit not incuding the jet data (outer error bands). The uncertainties are 
shown as fractional differences 
from the central values of the fits, for various values of $Q^2$. 
The total experimental uncertainty includes the statistical, uncorrelated 
and correlated systematic uncertainties and normalisations, for both fits.
} 
\label{glujets}
\end{figure}

In Fig.~\ref{glujets} the uncertainty of the gluon distribution for fits
with and without the jet data are compared. 
The shapes of the PDFs 
are not changed significantly by the addition of jet data, even though the 
gluon parameterisation is sufficiently flexible to allow this, indicating that
there is no tension between the jet data and the inclusive cross-section data. 
Although the jet data constrain the gluon mainly in the range  
$ 0.01 \leqsim \xi \leqsim 0.4$, 
the momentum sum-rule ensures that the indirect 
constraint of these data is still significant at higher $x$. 
The decrease in the uncertainty on 
the gluon distribution is striking; for example at $Q^2 = 7$~GeV$^2$ and 
$x = 0.06$ the uncertainty is reduced from $17\%$ to $10\%$. A similar decrease
in uncertainty by a factor of about two is found in this mid-$x$ range, over 
the full $Q^2$ range.

\begin{figure}[ht]
\begin{center}
~\epsfig{file=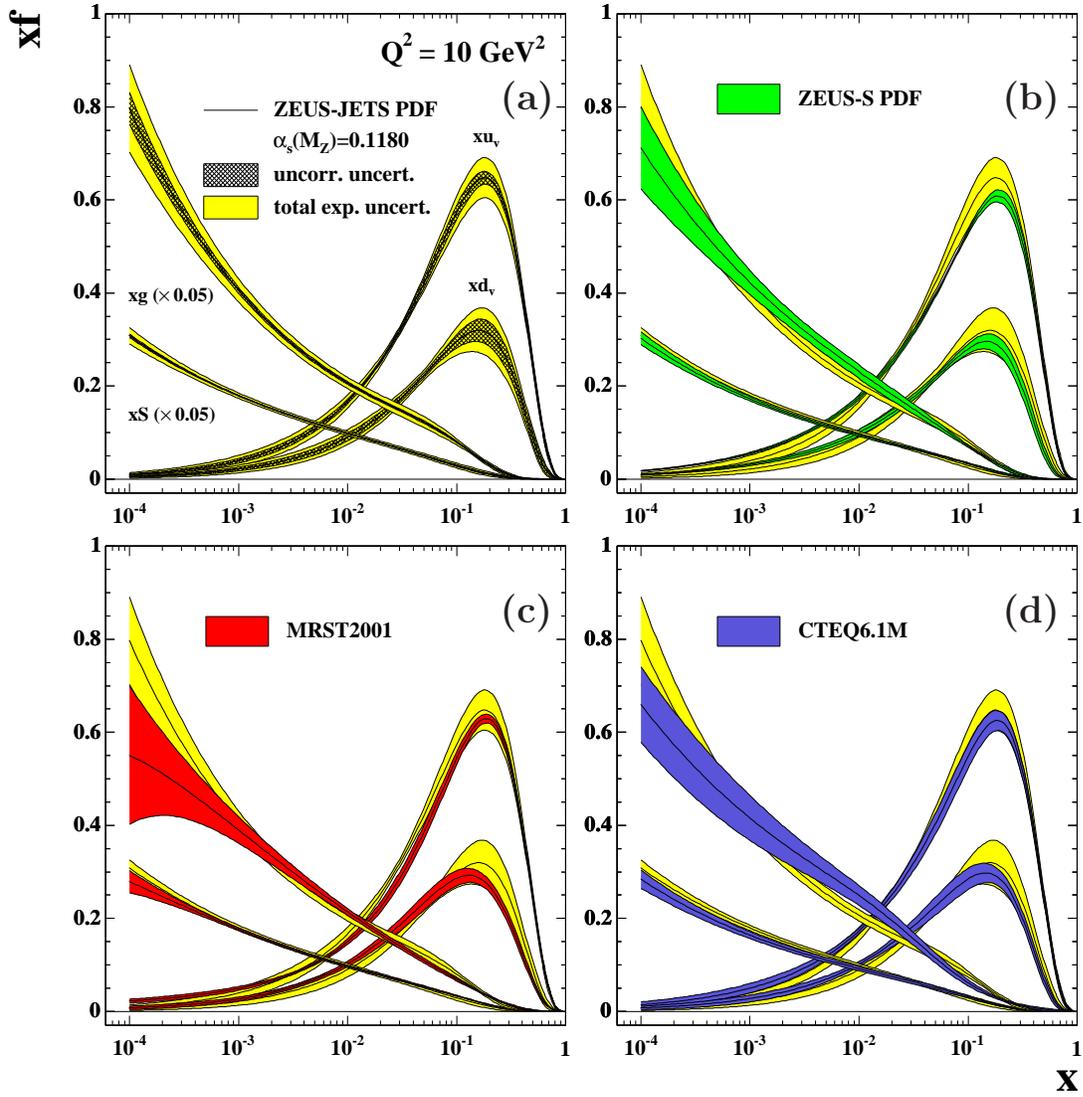,width=1.0\textwidth}
\put(-245,400){\makebox(0,0)[tl]{\large \bf (a)}} 
\put(-55,400){\makebox(0,0)[tl]{\large \bf (b)}} 
\put(-245,206){\makebox(0,0)[tl]{\large \bf (c)}} 
\put(-55,206){\makebox(0,0)[tl]{\large \bf (d)}} 
\end{center}
\caption{(a) PDFs extracted from the ZEUS-JETS fit. (b) PDFs extracted from 
the ZEUS-JETS fit compared to ZEUS-S PDFs. (c) PDFs extracted from the 
ZEUS-JETS fit compared to MRST2001 PDFs. (d) PDFs extracted from the 
ZEUS-JETS fit compared to CTEQ6.1 PDFs. 
The total experimental uncertainty bands are shown for each PDF set.}
\label{fig:summary}
\end{figure}

In Fig.~\ref{fig:summary}, 
the valence, sea and gluon PDFs 
are compared for the ZEUS-JETS fit and the previous ZEUS-S global PDF analysis.
There is good agreement between the ZEUS PDF extractions. The figure also 
compares the MRST
and CTEQ PDFs
to the ZEUS-JETS PDFs.
These PDFs are compatible with the ZEUS PDFs,
 considering the size of the uncertainties on each of the PDF sets.

\subsection{PDF Uncertainties}
\label{sec:uncertain}

\begin{figure}[ht]
\begin{center}
~\epsfig{file=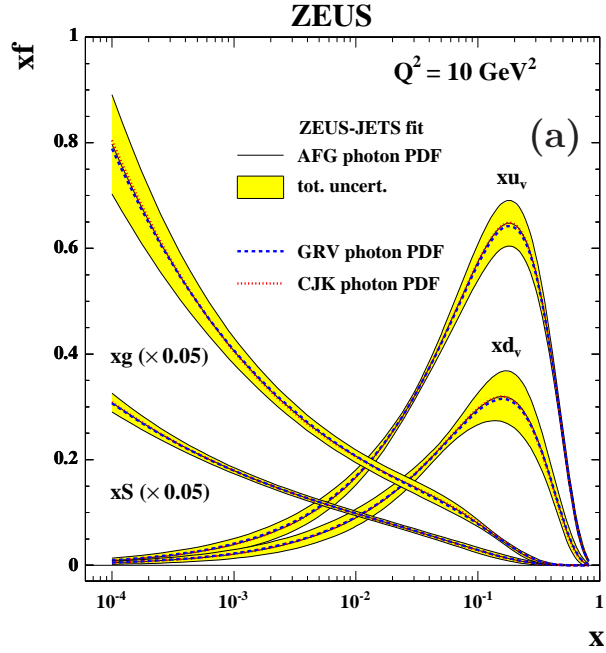,width=9cm}
\put(-45,206){\makebox(0,0)[tl]{\large \bf (a)}} \\
~\epsfig{file=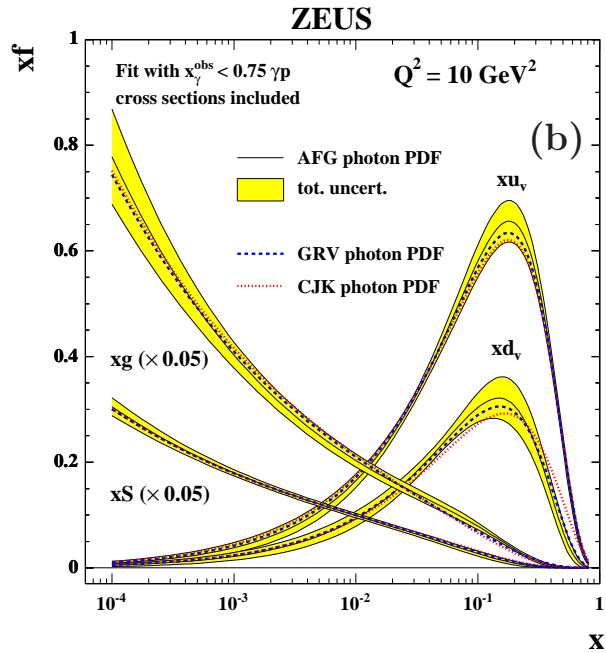,width=9cm}
\put(-45,206){\makebox(0,0)[tl]{\large \bf (b)}}
\end{center}
\caption{(a) PDFs extracted from the ZEUS-JETS fit using different photon 
PDFs. The AFG photon PDF is used to obtain the central line, the GRV photon PDF
gives the dashed line and the CJK photon PDF gives the dotted line. 
(b) PDFs extracted from a fit in which the resolved photoproduction 
cross-sections are included in addition to all the standard data sets for the
ZEUS-JETS fit. The AFG photon PDF is used to obtain the central line, 
the GRV photon PDF
gives the dashed line and the CJK photon PDF gives the dotted line. 
The total experimental 
error bands shown in these figures were obtained using the  
AFG photon PDF; for details see the caption to Fig.~\ref{glujets}.
 } 
\label{fig:resolved}
\end{figure}

The following sources of model uncertainty have been included in the PDF 
uncertainty bands:

\begin{itemize}
\item the value of $Q^2_0$ was varied 
in the range $4 < Q^2_0 < 10$~GeV$^2$;
\item the forms of the input PDF parameterisations were changed, 
by modifiying the form $(1 + p_4 x)$ to $(1 + p_4 x + p_5 \sqrt{x})$ 
for the valence parameterisations 
and by varying the choice of constraints applied to the sea parameterisation 
as explained in Section~\ref{sec:anal};
\item the standard $E_T$ cuts applied to the jet data 
were raised to $E_T^B > 10~\gev$ and $E_T^{\rm jet1} > 17~\gev$ for DIS 
jets and photoproduced jets, respectively, since  
there are some small discrepancies between the fit predictions and the jet data 
at the lowest
transverse energies\footnote{This is also the case for the MRST and 
CTEQ PDFs~\cite{pl:b547:164,epj:c23:615}.};
\item
the hadronisation corrections applied to the jet data have been varied by half 
the difference between the values obtained from the {\sc Herwig} and 
{\sc Pythia} MC programmes for the photoproduced jet 
cross sections~\cite{epj:c23:615} and by the 
variance of the values obtained from the {\sc Ariadne}, {\sc Lepto} and 
{\sc Herwig} MC
programmes for the DIS jet cross sections~\cite{pl:b547:164}. 
The uncertainties on the hadronisation corrections determined by these 
procedures are $< 1\%$; they lead to uncertainties in the PDFs which are 
small in comparison to the experimental uncertainties;
\item
as explained in Section~\ref{sec:jets}, the photoproduction data used in 
the fit are enriched with direct photon processes
by the cut $x^{\rm obs}_\gamma > 0.75$; however it is not possible to select 
jet cross sections that are completely independent of photon structure. 
Therefore
the sensitivity of the fit results to the input photon PDFs was investigated.
In Fig.~\ref{fig:resolved}a the proton PDFs
extracted from the ZEUS-JETS fit using the AFG photon 
PDFs~\cite{zfp:c64:621} are compared with those extracted using 
the GRV~\cite{pr:d45:3986,pr:d46:1973} and CJK~\cite{pr:d70:093004} 
photon PDFs.  
There is no visible difference in the extracted proton PDFs. 
In Fig.~\ref{fig:resolved}b this comparison is shown for 
a fit in which the `resolved' photon cross sections,  
$x^{\rm obs}_\gamma < 0.75$~\cite{epj:c23:615}, have been included.
A significant difference is now observed between the extracted
proton PDFs using the AFG, GRV, or CJK photon PDFs. 
Note that this difference
is greatest in the region of $x$ where the jet data have the most significant 
impact in reducing the uncertainty of the gluon PDF. Thus, 
although the 
addition of the resolved photoproduction cross sections reduces the 
experimental uncertainty on the extracted gluon PDF, it introduces a model 
uncertainty due to the limited knowledge of the photon PDFs
which outweighs this advantage. Hence the present analysis used only the
photoproduction cross sections with $x^{\rm obs}_\gamma > 0.75$.    
The difference in the proton PDFs 
extracted using the AFG and GRV photon PDFs was used to 
estimate the small residual model uncertainty due to the photon PDF in the 
ZEUS-JETS fit.
\end{itemize}

\begin{figure}[Htp] 
\vspace*{13pt}
\centerline{
\psfig{figure=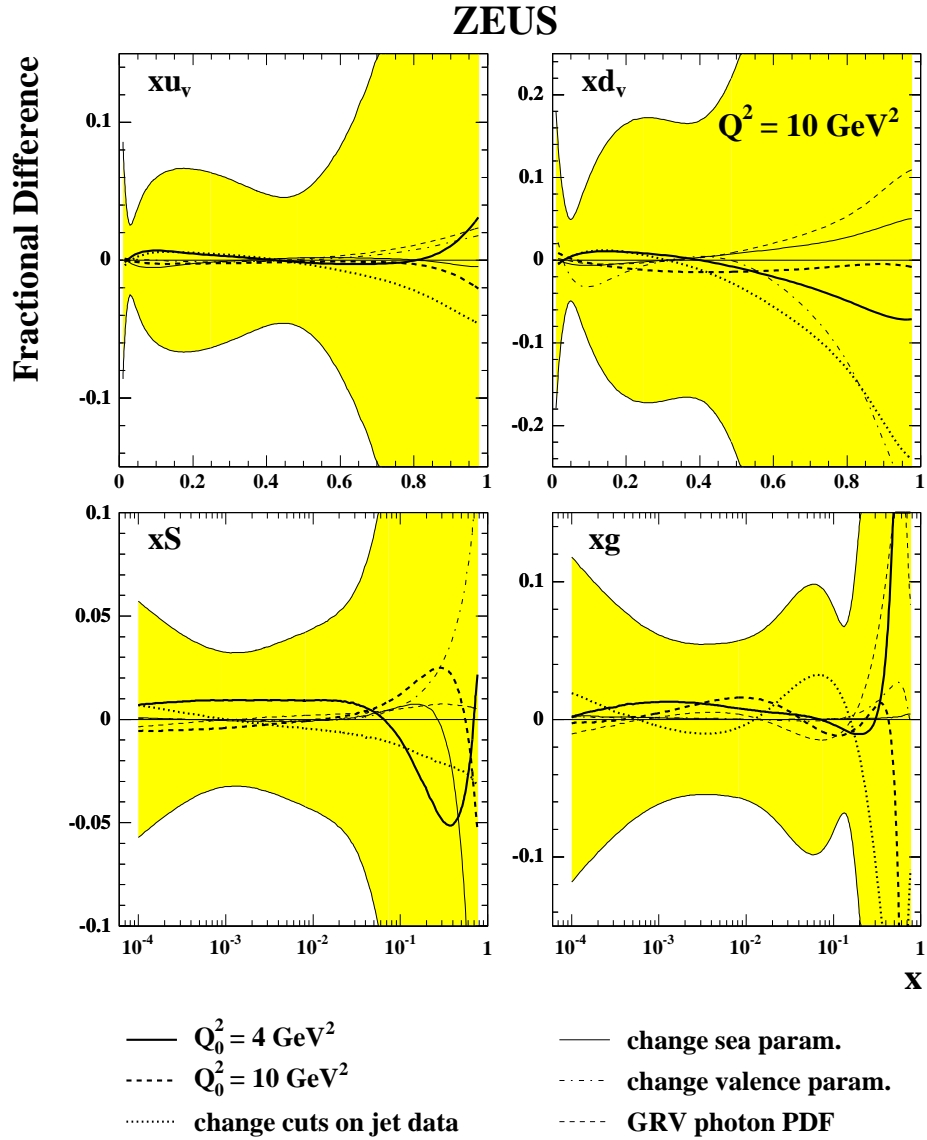,width=0.8\textwidth}}
\caption {Model variations discussed in the text are illustrated as
fractional differences from the ZEUS-JETS central value for all the PDFs.
For comparison, the shaded band shows the total experimental uncertainty.
}
\label{fig:models}
\end{figure}

The effect of some of the larger model variations listed above 
on the shapes of the extracted
PDFs is illustrated in Fig.~\ref{fig:models}. These model
variations are a much smaller source of uncertainty 
than the experimental uncertainties.

In addition to these model uncertainties a variety of cross-checks have been 
made:
\begin{itemize}
\item the minimum $x$ of data entering the fit was raised to 
$x > 5 \times 10^{-4}$, and the minimum $Q^2$ of data entering the fit was
 raised to $Q^2 > 4.5$~GeV$^2$. These variations did not produce any 
significant changes in the PDF parameters; 
\item 
the ZMVFN heavy quark production scheme wasused instead of the TRVFN
scheme. The jet data are all at sufficiently high scale that the TRVFN and 
ZMVFN schemes are equivalent. However, it is well known that 
the use of  the ZMVFN scheme makes small differences to the shape of 
the gluon at  $x < 10^{-3}$. This shift is well within the experimental 
uncertainty bands;
\item 
the choices of factorisation and renormalisation scale have been varied. 
The choice of $Q$ is not in dispute for inclusive 
DIS processes. However, the scale choices for jet-production are 
not so unambiguous. Thus, factorisation and renormalisation scales were
varied by a factor of $\surd{2}$
for both the DIS jets and the photoproduced jets and
additionally the conventional scale $\mu_F=\mu_R=Q$ 
for the inclusive cross-section 
data was varied by the same 
factor\footnote{Larger variations, by a factor of 2, 
are not presented since they produce fits with unacceptably large $\chi^2$.
The acceptability of a $\chi^2$ is judged by the hypothesis testing 
criterion~\cite{pr:d67:012007}
such that the variation from the minimum should not exceed $\sim\sqrt{2N}$, 
where $N$ is the number of degrees of freedom. In the ZEUS-JETS fit, 
$\sqrt{2N}=33$.}. 
The renormalisation scale for the 
DIS jet data was also changed from $\mu_R = E^B_T$ to $\mu_R = Q$.
These changes in the choice of 
scale produced changes in the shapes of the PDFs which are small in comparison 
to the experimental uncertainties; 
\item 
a Hessian fit was performed to the same data sets as for the ZEUS-JETS fit. 
The central values of the PDF parameters were found to be similar to those of 
the ZEUS-JETS fit, well within the latter's uncertainties. 
In the Hessian fitting method~\cite{jp:g28:2609,jp:g28:2717}, 
the theoretical prediction is used to 
determine the optimal correlated systematic shifts of the data.  
The correlated systematic uncertainties are assumed to be 
Gaussian distributed.  This assumption is not correct for the data sets 
considered here, and the resulting 
uncertainties of the fit are underestimated.  On the other hand, the method 
has a $\chi^2$ which is a well defined measure of the 
goodness-of-fit, not available in the Offset method.
The $\chi^2$ per degree of freedom of the Hessian fit was 1.12 for 566 
degrees of freedom\footnote{If the $E_T$ cuts applied to the jet data are 
raised, as described in Section~\ref{sec:uncertain}, 
the $\chi^2$ per degree of 
freedom of the Hessian fit becomes 1.01 for 554 degrees of freedom.}.

\end{itemize}

\begin{figure}[Htp] 
\vspace*{13pt}
\centerline{
\psfig{figure=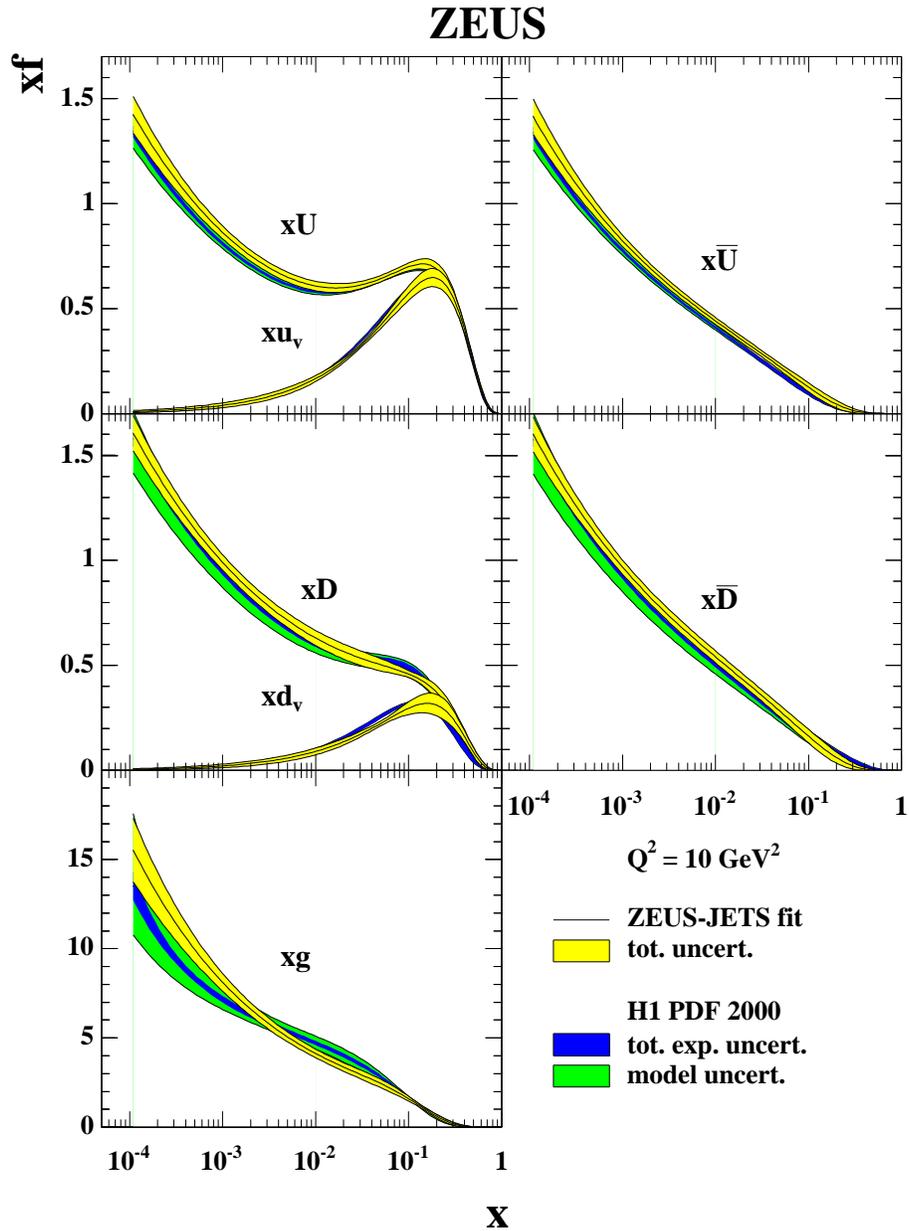,width=0.8\textwidth}
}
\caption {Comparison of the PDFs extracted from the ZEUS-JETS fit with those 
extracted in the H1 2000 PDF analysis.
For each analysis the 
total experimental error bands and the model error bands are included. 
However, the model uncertainty is not visible for the ZEUS-JETS fit.
}
\label{fig:H1}
\end{figure}

Figure~\ref{fig:H1} shows the ZEUS-JETS PDFs compared to those of the
H1 2000 PDF analysis~\cite{epj:c30:1}. 
The comparison is done in terms of the $xU = x(u + c)$, 
$x\bar{U}=x(\bar{u} + \bar{c})$, $xD = x(d + s)$, 
$x\bar{D} = x(\bar{d} + \bar{s})$ and gluon PDFs, 
which have been directly extracted by H1. The PDFs 
extracted by ZEUS and H1 are broadly compatible. 
Note that the Hessian method of 
treatment of the correlated systematic uncertainties used 
in the H1 fit results in a smaller 
experimental uncertainty on the gluon PDF~\cite{jp:g28:2609}, 
but the model uncertainty is significant. By contrast, 
the Offset method of treatment of correlated systematic uncertainties  
used in the ZEUS fit results in a larger experimental uncertainty,
so that it dominates in comparison to the model uncertainties.

\subsection{Comparison to Tevatron jet data}

It has been suggested that PDF 
fits to DIS data alone cannot produce a hard enough 
high-$x$ gluon to describe the high-$E_T$ inclusive 
jet cross sections measured at the Tevatron~\cite{epj:c23:73}.
 To investigate this issue,
the ZEUS-JETS PDFs were used to make predictions for the CDF 
jet cross sections. 
The information on the correlated systematic 
uncertainties of the CDF data is supplied in such a way that it is possible to 
make a fit 
to these data by the Hessian method. In such a fit, 
the PDF parameters are fixed
but the eight systematic uncertainties are freed.  
The $\chi^2$ of the CDF jet data with respect to the ZEUS-JETS fit was 
calculated using this procedure and $\chi^2 = 48.9$ was obtained. This is to be
compared to $\chi^2 = 46.8$ which was obtained by the CDF 
collaboration~\cite{pr:d64:032001}, 
using the same procedure, for a fit to the CTEQ4HJ PDFs, which were specially 
developed to fit the CDF jet data. Thus, the ZEUS-JETS PDFs give an acceptable
description of the CDF jet data.

\section{Extraction of $\alpha_s$}
\label{sec:alphas}

The strong 
correlation between the gluon shape and the 
value of $\alpha_s(M_Z)$, which affects fits to inclusive cross-section data 
alone, can be broken by including the
jet production cross-section data, which are dependent on the gluon PDF and the
value $\alpha_s(M_Z)$ in a different way from the total cross section. 
Jet production cross sections are directly dependent on the 
gluon PDF through the BGF process, but jet production also proceeds though the 
QCDC process,  which dominates the cross section at large scales. This process 
depends on $\alpha_s(M_Z)$ and the quark densities, which are directly 
determined from the inclusive cross-section data. 
Thus the addition of jet data allows an extraction of
$\alpha_s(M_Z)$ that is not strongly correlated to the shape of the gluon
PDF.

In previous determinations of $\alpha_s(M_Z)$ using ZEUS jet 
data~\cite{pl:b547:164,pl:b507:70,epj:c31:149,pl:b558:41,np:b700:3,pl:b560:7},
the uncertainty from the correlation to the PDFs was taken into account 
by using
PDFs from the global fits of CTEQ and MRST, which were determined assuming 
different values of
$\alpha_s(M_Z)$. In the present analysis this correlation 
is directly included by fitting the PDF parameters and $\alpha_s(M_Z)$ 
simultaneously. The conditions for the ZEUS-JETS-$\alpha_s$ fit are otherwise 
the same as for the ZEUS-JETS fit. The value
\[ \alpha_s(M_Z) = 0.1183 \pm 0.0007({\rm uncorr.}) \pm 0.0022({\rm corr.}) 
\pm 0.0016({\rm norm.}) \pm 0.0008({\rm model})
\]
was obtained, 
where the four uncertainties arise from the following sources: statistical
 and other uncorrelated sources; experimental correlated systematic sources
excluding normalisation uncertainties; normalisation uncertainties; and 
model uncertainty. Here the uncertainty on $\alpha_s(M_Z)$, 
which usually comes from the correlation to the PDF shapes, is 
automatically included in the experimental uncertainties. 
The sources of model uncertainty were
discussed in Section~\ref{sec:uncertain}. In addition to the model 
uncertainties included in the PDF extraction, the following extra sources have
been included in the model uncertainty for $\alpha_s(M_Z)$: variation of the
$Q^2$ and $x$ cuts on the data, as specified in 
Section~\ref{sec:uncertain}; and the use of the ZMVFN instead of the RTVFN 
scheme for heavy quark production. 

This extraction is at NLO. A crude estimate of the effect of terms beyond 
NLO can be made by variation of the choice of $\mu_R$. This scale was
varied by a factor of $\surd{2}$ for all the data sets entering into the fit, 
as described in Section~\ref{sec:uncertain}.  
The most significant effect comes from the 
variation of the renormalisation scale for the photoproduction process. 
These scale changes
produced shifts of $\Delta\alpha_s(M_Z) \sim \pm 0.005$. 
   
\begin{figure}[Hbp] 
\vspace*{13pt}
\centerline{
\psfig{figure=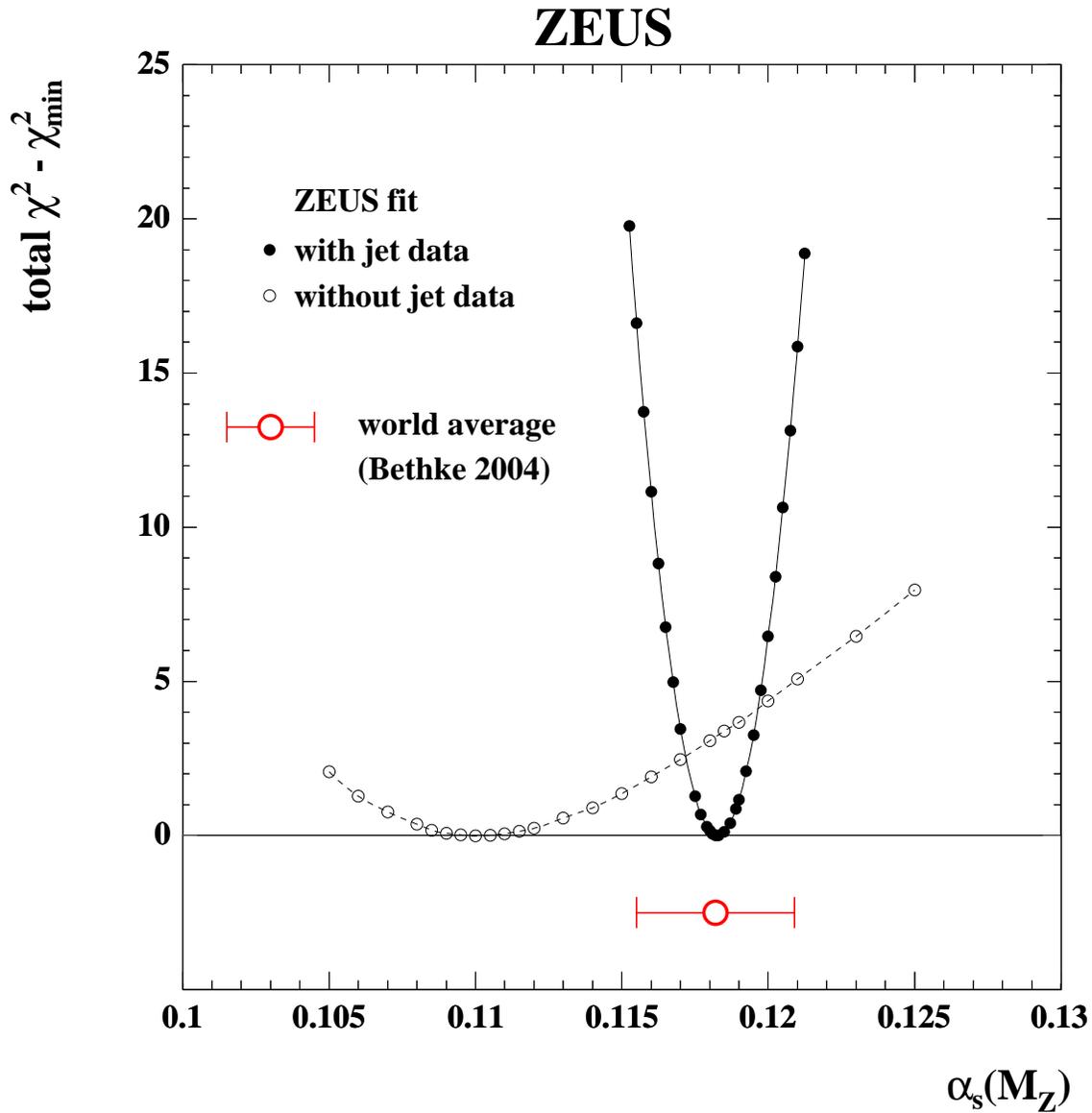,width=1.0\textwidth}}
\caption {The $\chi^2$ profile as a function of $\alpha_s(M_Z)$ for the 
ZEUS-JETS-$\alpha_s$ fit (black dots) and for a similar fit not including the jet data 
(clear dots). The ordinate is given in terms of the difference between the
total $\chi^2$ and the minimum $\chi^2$, for each fit.}
\label{fig:chiprof}
\end{figure}

Figure~\ref{fig:chiprof} illustrates that 
the improved accuracy of the extraction 
of $\alpha_s(M_Z)$ in the ZEUS-JETS-$\alpha_s$ fit is 
due to the inclusion of the jet data. The $\chi^2$ 
profile around the minimum is shown as a function of $\alpha_s(M_Z)$ for the 
ZEUS-JETS-$\alpha_s$ fit and a similar fit in which the jet data are not 
included. The value of $\alpha_s(M_Z)$ extracted is in agreement with recent 
determinations using measurements in 
DIS~\cite{pr:d67:012007,epj:c21:33,pl:b547:164,pl:b507:70,epj:c31:149,pl:b558:41,np:b700:3,epj:c19:289} 
and photoproduction of jets~\cite{pl:b560:7} at HERA and with the current 
world average of $0.1182 \pm 0.0027$~\cite{jp:g26:r27,hep-ex-0407021}.

\begin{figure}[Hbp] 
\vspace*{13pt}
\centerline{
\psfig{figure=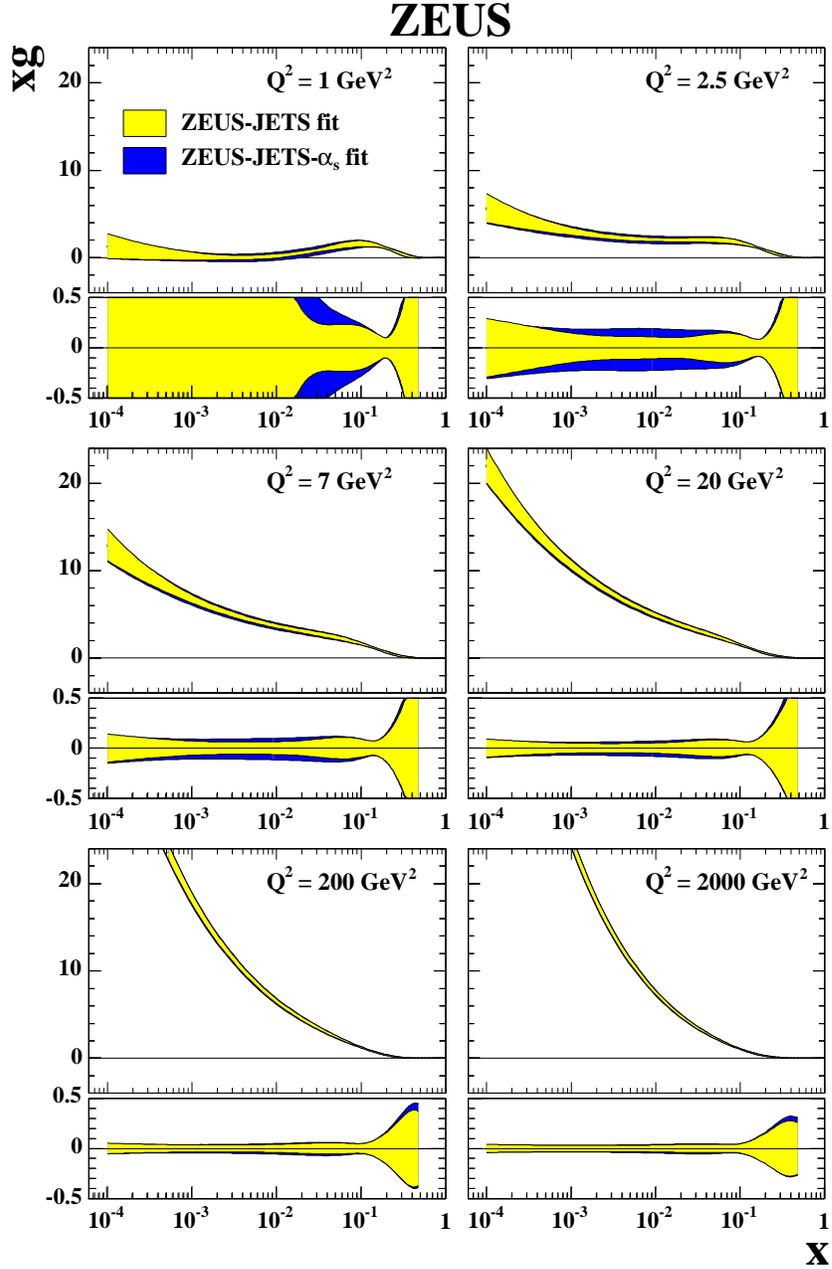,width=0.7\textwidth}}
\caption {Gluon distributions extracted from the ZEUS-JETS-$\alpha_s$ fit.
The uncertainties on these distributions are shown beneath each distribution 
as fractional differences from the central value.
 The inner error bands show the total uncertainty including statistical, 
uncorrelated and correlated
experimental systematic uncertainties, 
normalisations and model uncertainties and the outer error 
bands show the additional
uncertainty in the gluon coming from the variation 
of $\alpha_s(M_Z)$.}
\label{fig:gluealf}
\end{figure}

The extracted value of 
$\alpha_s(M_Z)$ is close to the fixed value used in the ZEUS-JETS fit, 
and there are therefore no significant changes in the central values of the 
PDF parameters. The uncertainties of the valence and sea PDFs are also
unaffected.
However, there is some increase in the overall uncertainty of the gluon PDF 
because a weak correlation remains between $\alpha_s(M_Z)$ and the 
gluon PDF parameters. This is illustrated for various $Q^2$ values
in Fig.~\ref{fig:gluealf}. 

The input of the jet data results in a much reduced uncertainty on the
extracted value of $\alpha_s(M_Z)$ compared to the previous
ZEUS-$\alpha_s$ analysis~\cite{pr:d67:012007}. 
Since the present analysis was performed 
within a single experiment, the contributions from normalisation uncertainties 
and from correlated systematic uncertainties are both significantly reduced. 
In consequence, the total uncertainty on the gluon, 
including the uncertainty due to $\alpha_s(M_Z)$, is reduced in comparison 
to the total gluon uncertainty determined in the ZEUS-$\alpha_s$ global fit.

\section{Summary}
\label{sec:summ}

Due to the precision and kinematic coverage of the ZEUS data, it is now 
possible to extract proton PDFs and $\alpha_s(M_Z)$ in a fit to data from a 
single experiment with minimal external input. The
ZEUS high-$Q^2$ cross sections were used to constrain the valence PDFs,
 ZEUS low-$Q^2$ NC data were used to constrain the low-$x$ sea and gluon
distributions and ZEUS data on jet production were used to
constrain the mid- to high-$x$ gluon. This provides a compelling demonstration 
of QCD factorisation, showing that NLO QCD in the framework of the Standard 
Model is able to simultaneously describe 
inclusive cross sections and jet cross sections.
The additional constraint on the gluon PDF from the 
jet production data allows an accurate extraction of the value of 
$\alpha_s(M_Z)$ in NLO QCD, 
\[ \alpha_s(M_Z) = 0.1183 \pm 0.0028({\rm exp.}) \pm 0.0008({\rm model}) .
\]
The uncertainty in $\alpha_s(M_Z)$ due to terms beyond NLO has been
estimated as  $\Delta\alpha_s(M_Z) \sim \pm 0.005$, by variation of the 
choice of scales. This is the first extraction of $\alpha_s(M_Z)$ from HERA 
data alone.

The total uncertainty on the gluon PDF is reduced in comparison to the
ZEUS-$\alpha_s$ global fit because of the greater precision of 
the $\alpha_s(M_Z)$ 
measurement. The uncertainties on the valence PDFs are becoming
competitive with those of the global fits, and they are not subject to 
uncertainties from heavy-target corrections, higher-twist contributions or 
isospin-symmetry assumptions. The precision of PDFs extracted
from the global fits is now limited by the systematic uncertainties of the 
contributing experiments, whereas the precision of the present fit using
ZEUS data only is limited by the statistical uncertainties and so 
further improvement can be expected
when higher precision HERA-II data become available.

 \section*{Acknowledgements} 
We thank the DESY directorate for their strong support and encouragememt. 
The remarkable achievements of the HERA machine group were essential for the 
successful completion of this work and are greatly appreciated. The design,
construction and installation of the ZEUS detector has been made possible by 
the effort of many people who are not listed as authors. We acknowledge 
helpful discussions with Robert Thorne and Dick Roberts.

\providecommand{\etal}{et al.\xspace}
\providecommand{\coll}{Collab.\xspace}
\catcode`\@=11
\def\@bibitem#1{%
\ifmc@bstsupport
  \mc@iftail{#1}%
    {;\newline\ignorespaces}%
    {\ifmc@first\else.\fi\orig@bibitem{#1}}
  \mc@firstfalse
\else
  \mc@iftail{#1}%
    {\ignorespaces}%
    {\orig@bibitem{#1}}%
\fi}%
\catcode`\@=12

\newpage
\begin{table}
\begin{center}
\begin{tabular}{|l|c|c|c|c|}
\hline
\footnotesize{Data Set} & \footnotesize{Ndata} & \footnotesize{Norm} & \footnotesize{Nsys}  & \footnotesize{Kinematic range}   \\
           &   &         &  & \footnotesize{of the data} \\
\hline
\footnotesize{NC $e^+p$ 96-97~\cite{epj:c21:443}} & 242   & $2\%$ & 10 & 
\footnotesize{$2.7 < Q^2 < 30,000$~GeV$^2$}\\
 & &($1\%$) &1,2,3,4,5,6,7,8,9,10 & \footnotesize{$6.3 \times 10^{-5} < x < 0.65$}\\
\hline
\footnotesize{CC $e^+p$ 94-97~\cite{epj:c12:411}} & 29   & $2\%$ &  3 &
\footnotesize{$280 < Q^2 < 17,000$~GeV$^2$}\\
 & & &5,6,11 &\footnotesize{$0.015 < x < 0.42$}\\
\hline
\footnotesize{NC $e^-p$ 98-99~\cite{epj:c28:175}} & 92   & $1.8\%$ &  6  &
\footnotesize{$200 < Q^2 < 30,000$~GeV$^2$}\\
 & & &12,13,14,15,16,11 &\footnotesize{$0.005 < x < 0.65$}\\
\hline
\footnotesize{CC $e^-p$ 98-99~\cite{pl:b539:197}} & 26   & $1.8\%$ & 3   &
\footnotesize{$280 < Q^2 < 17,000$~GeV$^2$}\\
 & & &17,18,11 &\footnotesize{$0.015 < x < 0.42$}\\
\hline
\footnotesize{NC $e^+p$ 99-00~\cite{pr:d70:052001}} & 90   & $2\%$ &  8 &
\footnotesize{$200 < Q^2 < 30,000$~GeV$^2$}\\
 & & &12,13,14,15,19,11,20,21 &\footnotesize{$0.005 < x < 0.65$}\\
\hline
\footnotesize{CC $e^+p$ 99-00~\cite{epj:c32:16}}  & 30   & $2\%$ &  3 & 
\footnotesize{$280 < Q^2 < 17,000$~GeV$^2$}\\
 & & &17,-18,11 &\footnotesize{$0.008 < x < 0.42$}\\
\hline
\footnotesize{DIS jets $e^+p$ 96-97~\cite{pl:b547:164}}  & 30   & $2\%$ & 1 & 
\footnotesize{$125 < Q^2 < 30,000$~GeV$^2$}\\
 & & &22 &\footnotesize{$8 < E^B_T  < 100$ GeV}\\
\hline
\footnotesize{$\gamma p$ dijets 96-97~\cite{epj:c23:615}}  &  38   & $2\%$ & 1 &
\footnotesize{$14 < E_T^{\rm jet1}  < 75$ GeV}\\
$x^{\rm obs}_\gamma > 0.75$ & & &22 &\\
\hline
\end{tabular}

\caption{The number of data points (Ndata), normalisation uncertainties (Norm)
and number of point-to-point correlated systematic uncertainties (Nsys) 
are detailed for each of the data sets used in the ZEUS-JETS 
fit. The kinematic regions of the data sets are also given.
The number of independent correlated systematic uncertainties is specified 
as follows. Each independent source of uncertainty is assigned a number 
in the order of the systematic uncertainties as 
given in the corresponding publication. These numbers are given in the column
headed Nsys, for each data set. For example, for the CC $e^+p$ 94-97 
data set, the first two systematic uncertainties are fully correlated to 
the fifth and sixth systematic uncertainties for the NC $e^+p$ 96-97 data set.
Note also that the second systematic uncertainty for the CC $e^+p$ 99-00 data 
set is fully anti-correlated to the second systematic uncertainty for the 
CC $e^-p$ 98-99 data.
The normalisation uncertainties are applied as follows. There are two 
normalisation uncertainties for the NC $e^+p$ 96-97 data: 
an overall uncertainty and the relative uncertainty (indicated in parentheses) 
of the data with $Q^2 < 30$GeV$^2$, with respect to the higher $Q^2$ data.
The CC $e^+p$ 94-97 data are dominated by the 96-97 data, so that the same 
overall normalisation uncertainty is applied to this data set. The two jet 
production data sets also share the overall normalisation uncertainty of the 
96/97 data. The NC and CC $e^-p$ 98-99 data share a common normalisation 
uncertainty as do the  NC and CC $e^+p$ 99-00 data.} 

\label{tab:chisq}
\end{center}
\end{table}

\clearpage
\begin{table}
\begin{center}
\vspace{3.0cm}
\begin{tabular}{|l|c|c|c|c|}
\hline
PDF & $p_1$& $p_2$ & $p_3$ & $p_4$ \\
\hline
 $xu_v$ &    \footnotesize{($3.1 \pm 0.7 \pm 1.2$)} &
             \footnotesize{$0.64 \pm 0.05 \pm 0.08$}                      &
             \footnotesize{$4.06 \pm 0.18 \pm 0.24$}   &
             \footnotesize{$2.3 \pm 1.1 \pm 1.0$}   \\
\hline
 $xd_v$ &
             \footnotesize{($1.7 \pm 0.3  \pm 0.5$} &
             \footnotesize{$0.63 \pm 0.05 \pm 0.08$}                      &
             \footnotesize{$4.8 \pm 0.7 \pm 1.0$}   &
             \footnotesize{$2.6 \pm 2.2 \pm 2.3$}      \\
\hline
 $xS$ &      \footnotesize{$0.72 \pm 0.03 \pm 0.10$} &
             \footnotesize{$-0.217 \pm 0.005 \pm 0.020$}&
             \footnotesize{$ 7.0 \pm 0.8 \pm 2.0$}     &
              \footnotesize{$0$}     \\
\hline
 $xg$  &     \footnotesize{($0.9 \pm 0.1 \pm 0.3$)} &
             \footnotesize{$-0.28 \pm 0.02 \pm 0.04$}  &
             \footnotesize{$10.2 \pm 0.7 \pm 2.1$}      &
             \footnotesize{$16 \pm 4 \pm 10$}                        \\

\hline
\end{tabular}
\caption{Table of PDF parameters at $Q^2_0 = 7$~GeV$^2$, as determined from
the ZEUS-JETS fit. The first uncertainty given originates from statistical
and other uncorrelated sources and the second uncertainty is the
additional
contribution from correlated systematic uncertainties.
The numbers in parentheses were derived from the fitted parameters via 
the number and momentum sum-rules.
}
\label{tab:param}
\end{center}
\end{table}

\

\clearpage

\clearpage

\clearpage

\clearpage

\clearpage

\clearpage

\clearpage

\clearpage

\clearpage

\clearpage

\clearpage

\clearpage

\clearpage

\clearpage

%
%
\end{document}